\documentclass{aa} 
\usepackage{graphicx}
\usepackage[varg]{txfonts}
\usepackage{natbib}
\usepackage{txfonts}
\usepackage[dvipsnames]{xcolor}
\usepackage[colorlinks=true, linkcolor={blue!70!black}, citecolor={blue!70!black}, urlcolor={blue!70!black}]{hyperref}

\usepackage{CJKutf8}
\begin{document} 

   \title{\textit{JWST} discovers an AGN ionization cone but only weak radiative-driven feedback in a powerful $z$$\approx$3.5 radio-loud AGN}
   \titlerunning{\textit{JWST} discovers weak AGN radiative-driven feedback in a powerful $z$$\approx$3.5 radio-loud AGN}


\author{Wuji Wang (\begin{CJK*}{UTF8}{gbsn}王无忌\end{CJK*})  \inst{\ref{inst1}}
        \and
        Dominika Wylezalek\inst{\ref{inst1}}
        \and 
        Carlos De Breuck\inst{\ref{inst2}}
        \and
        Jo\"{e}l Vernet\inst{\ref{inst2}}
        \and
        David S. N. Rupke\inst{\ref{instDR}}
        \and 
        Nadia L. Zakamska\inst{\ref{instJHU}}
        \and
        Andrey Vayner\inst{\ref{instJHU}}
        \and
        Matthew D. Lehnert\inst{\ref{insML}}
        \and 
        Nicole P. H. Nesvadba\inst{\ref{instNN}}
        \and
        Daniel Stern\inst{\ref{instDS}}
        }

    \institute{   Astronomisches Rechen-Institut, Zentrum f\"{u}r Astronomie der Universit\"{a}t Heidelberg, M\"{o}nchhofstr. 12-14, D-69120 Heidelberg, Germany\label{inst1}\\
                \email{wuji.wang@uni-heidelberg.de}$;\,\,$\email{wuji.wang\_astro@outlook.com}
   \and
   European Southern Observatory, Karl-Schwarzchild-Str. 2, D-85748 Garching, Germany\label{inst2}
   \and
    Department of Physics, Rhodes College, 2000 N. Parkway, Memphis, TN 38112, USA\label{instDR}
   \and
   Department of Physics and Astronomy, Bloomberg Center, Johns Hopkins University, 3400 N. Charles Street, Baltimore, MD 21218, USA\label{instJHU}
   \and
   Univ. Lyon, Univ. Lyon1, ENS de Lyon, CNRS, Centre de Recherche Astrophysique de Lyon UMR5574, 69230 Saint-Genis-Laval, France\label{insML}
   \and
   Universit\'{e} de la C\^{o}te d’Azur, Observatoire de la C\^{o}te d’Azur, CNRS, Laboratoire Lagrange, Bd de l’Observatoire, CS 34229, F-06304 Nice cedex 4, France\label{instNN}
   \and
   Jet Propulsion Laboratory, California Institute of Technology, 4800 Oak Grove
Drive, Pasadena, CA91109, USA\label{instDS}
        }
   \date{Received xxxx; Accepted 13 January 2024}
\authorrunning{Wuji Wang et al.}
 
  \abstract{We present the first results from a \textit{JWST} program studying the role played by powerful radio jets in the evolution of the most massive galaxies at the onset of Cosmic Noon. Using NIRSpec integral field spectroscopy, we detect 24 rest-frame optical emission lines from the $z=3.5892$ radio galaxy 4C+19.71. 4C+19.71 contains one of the most energetic radio jets known, making it perfect for testing radio-mode feedback on the interstellar medium (ISM) of a $M_{\star}\sim10^{11}\,\rm M_{\odot}$ galaxy. The rich spectrum enables line ratio diagnostics showing that the radiation from the active galactic nucleus (AGN) dominates the ionization of the entire ISM out to at least 25~kpc, the edge of the detection. Sub-kpc resolution reveals filamentary structures and emission blobs in the warm ionized ISM distributed on scales of $\sim5$ to $\sim20\,$kpc. A large fraction of the extended gaseous nebula is located near the systemic velocity. This nebula may thus be the patchy ISM which is illuminated by the AGN after the passage of the jet. A radiatively-driven outflow is observed within $\sim5\,$kpc from the nucleus. The inefficient coupling ($\lesssim 10^{-4}$) between this outflow and the quasar and the lack of extreme gas motions on galactic scales are inconsistent with other high-$z$ powerful quasars. Combining our data with ground-based studies, we conclude that only a minor fraction of the feedback processes is happening on $<25$kpc scales.
  } 


   \maketitle

%

\section{Introduction}

Active galactic nucleus (AGN) feedback can be categorized into either quasar-mode, where the output from the supermassive black hole (SMBH) is coupled radiatively to the gas, or radio-mode, where the feedback is due to the kinetic energy of the jet \citep[e.g.,][]{Fabian_2012,Harrison_2017}. The quasar-mode is known to be prevalent \citep[e.g.,][]{Harrison_2014}. The radio-mode also plays an important role in AGN--gas interactions, leading to quenching. This is especially true in massive galaxies in the early Universe, as shown by both numerical simulations and observations \citep[e.g.,][]{Heckman_2014,Mukherjee_2018,Kondapally_2023}. Indeed, despite the relatively short lifetime of a powerful jet ($\sim 10^{7}$ yr), observations indicate that it can still significantly impact the host galaxy \citep[e.g., total energy output by the jet $\gtrsim 10^{60}\,\mathrm{erg}$,][]{Nesvadba_2006a,Miley_2008}. Relativistic jets trigger shocks and drive outflows through an expanding over-pressured bubble \citep[e.g.,][]{Begelman_1989}. While hydrodynamic simulations demonstrated that jets have the ability to interact with the interstellar medium (ISM) out to several kiloparsecs \citep[e.g., creating turbulence, driving outflows and compressing the gas,][]{Dugan_2017,Mukherjee_2021}, observations paint a more complicated picture. Specially, jet--ISM interactions are found to be different between AGN with intermediate radio power \citep[$L_{\rm 1.4\, GHz}=10^{23-25}\,\mathrm{W\,Hz^{-1}}$, e.g.,][]{Mullaney_2013} and those with higher power \citep[e.g.,][]{Mukherjee_2016}. Moreover, it is difficult to study the feedback mechanisms of the most powerful jetted AGN due to observational limitations and the simultaneous presence of energetic, quasar-mode feedback. 

High-redshift radio galaxies (HzRGs) are the best targets to test AGN feedback in massive host galaxies \citep[$\sim10^{11}\,\mathrm{M_{\sun}}$,][]{DeBreuck_2010} because the gas, dust and stellar populations in their host galaxies can be observed without contamination from the bright, point-like quasar light. In contrast to low-$z$, where radio-mode AGN are believed to have low black hole accretion rates with inefficient radiative luminosity, HzRGs are observed to have both vigorous radio-mode and quasar-mode energy output \citep[e.g.,][]{vernet_2001,Nesvadba_2007a,Nesvadba_2017a,Nesvadba_2017b}. There are numerous studies focused on the outflows on several to 10s of kpc scale for both low- and high-redshift radio galaxies, which find evidence of them being jet-driven \citep[][]{Tadhunter_2001,Tadhunter_2007,Nesvadba_2006a,Nesvadba_2007a}. A closer look (scales of tens of pc) at the feedback in the vicinity of radio AGN is possible at low-$z$ \citep[e.g.,][]{Tadhunter_2003}. However, spotting the warm, ionized quasar outflow around Cosmic Noon \citep[][]{MandD_2014} down to sub-kpc scales is challenging. Therefore, it is still unclear how energetic jet-mode and quasar-mode feedback couple with the ISM, how efficient  these mechanisms are in driving outflows and on which scales the different feedback mechanisms dominate. All these are critical issues to be addressed before we can achieve a deeper understanding of quenching of star formation in these massive host galaxies \citep[][]{Falkendal_2019}.

We can now finally move a step forward thanks to the Near Infrared Spectrograph \citep[NIRSpec;][]{Jakobsen_2022} onboard \textit{JWST}, which provides an order of magnitude improvement in sensitivity. At least for HzRGs, there is evidence that the morphology and/or kinematics of the ionized ISM and even CGM (circumgalactic medium) gas are impacted by jets \citep[][]{Nesvadba_2008,Nesvadba_2017b,Falkendal_2021,wang2023}, which may indicate their role in feedback. However, the situation may be different when using deeper observations \citep[e.g.,][]{Wylezalek_2017}. For example, how does the ISM of HzRGs look in the vicinity of and inside the host galaxies (e.g., $\sim 10$ to 20 kpc)? NIRSpec integral field unit (IFU) observations of powerful quasars at Cosmic Noon have already revealed the detailed ionization mechanisms and quantified the outflow rates on galactic scales \citep[e.g.,][]{Wylezalek_2022q3d,Vayner_2023Q3D,Vayner_2023Q3D_2}. NIRSpec will revolutionize our view of the inner ISM of HzRGs where the detailed physics are still unclear. A resolution-matched comparison can finally be done for high- and low-$z$ radio AGN. NIRSpec IFU will also enable observational tests of simulated jet feedback \citep[e.g.,][]{Mukherjee_2016} in the early Universe. 

In this work, we present the first \textit{JWST} view of the ionized ISM of a $z\sim3.5$ radio-loud AGN. In Section \ref{sec:data_process}, we describe the observations and data processing. We then show in Section \ref{sec:results} the morphology of the warm ionized gas and identification of the observed optical emission lines from the extracted spectra. Finally, we discuss the impact of the jet on the ISM using line-ratio diagnostics and summarize in Section \ref{sec:summary}. Throughout this paper, we assume a flat $\Lambda$ cosmology with $H_{0} = 70\, \rm{km\,s^{-1}\,Mpc^{-1}}$ and $\Omega_{m}=0.3$. Following this cosmology, $\rm{1\,arcsec=7.25\, kpc}$ at the redshift of 4C+19.71. Throughout this work, we use $z=3.5892\pm0.0004$ as determined from atomic carbon, $[\ion{C}{i}]$(1-0) ($\nu_{\rm rest}=492.16\,\rm GHz$), as the systemic redshift \citep[][]{Falkendal_2021,kolwa2023}. Due to the relatively large synthesized beam \citep[$1.8\arcsec\times1.9\arcsec$,][]{Falkendal_2021} of the ALMA [\ion{C}{i}](1-0) observation, it is uncertain whether the molecular gas is located in the host galaxy (Appendix \ref{app:sys_z}). Though \textit{JWST} observes in a vacuum, we label the emission lines with their air wavelengths in $\mathrm{\mathring{A}}$ following the convention, for example [\ion{O}{iii}]5007. 

\begin{figure*}
    \centering
    \includegraphics[width=\textwidth,clip]{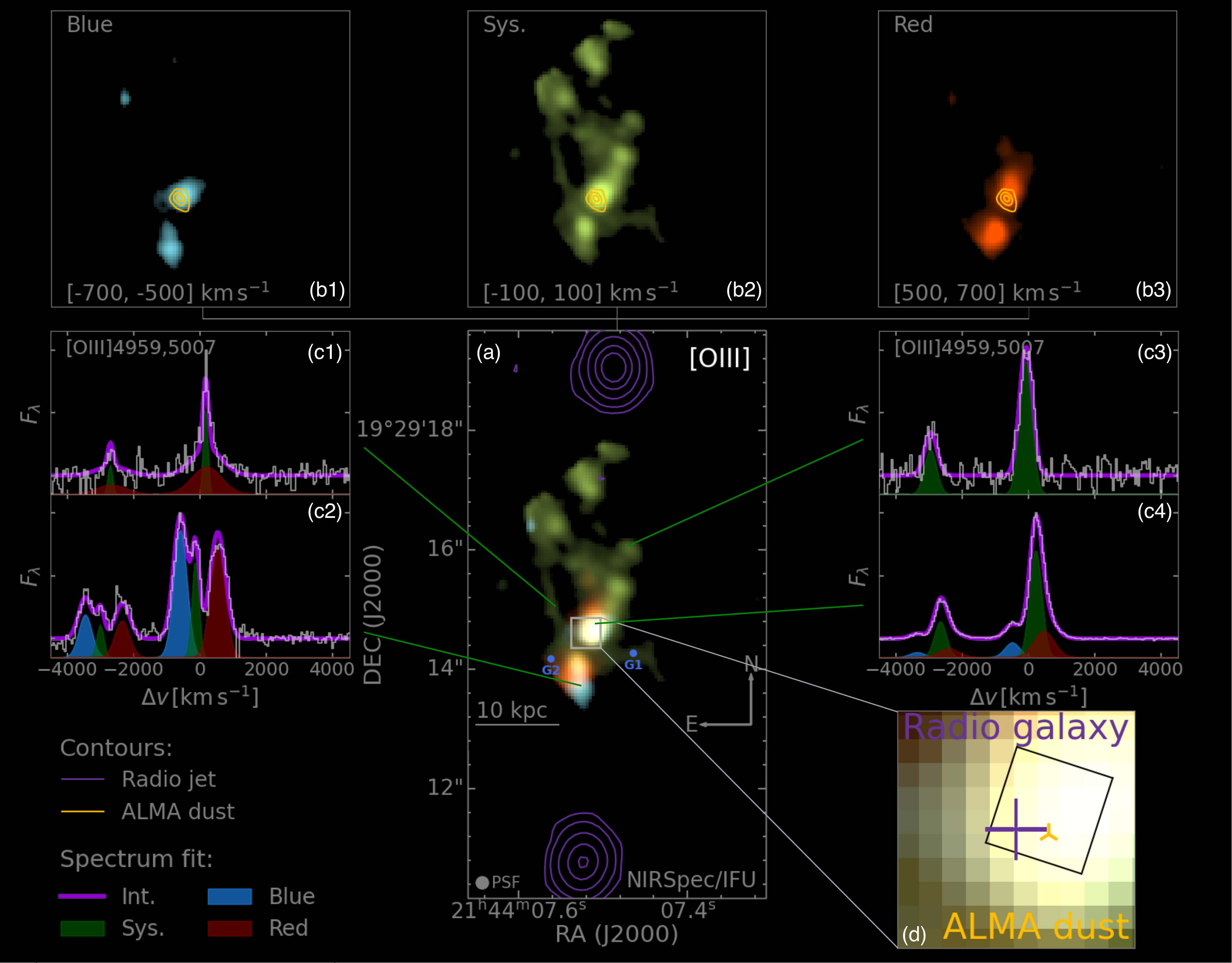}
    \caption{\textbf{(a)} Three-color composite image of narrow-band [\ion{O}{iii}]5007. We overlay the VLA 4.7\,GHz jet hot spots in dark purple contours. Blue dots indicate the position of foreground galaxies (see text). FWHM of the PSF is marked at the left corner. \textbf{(b)} ``Monochrome" narrow-band images of [\ion{O}{iii}]5007 collapsed at blue wing (b1), systemic redshift (b2) and red wing (b3), respectively. Yellow contours show dust continuum emission from the ALMA data. \textbf{(c)} Example spectra of continuum-subtracted [\ion{O}{iii}]$4959,5007$ extracted at four different spatial locations normalized to their peak flux density. We fitted the spectra using \texttt{q3dfit} with up to three kinematic components. The light purple curve indicates the overall fit. The individual components are shown in blue, green and red with a negative offset from the zero level. \textbf{(d)} Zoom-in view of the central $0.5\arcsec\times 0.5\arcsec$ region of the narrow-band image (gray box in panel a). Dark purple plus $+$ sign marks the position of the continuum emission of the radio galaxy determined from the NIRSpec cube while the yellow triple-spike triangle shows peak position of the ALMA 400.3\,GHz dust. The sizes of the markers represent the position uncertainty, $1.2\arcsec$ and $0.04\arcsec$ for radio galaxy and ALMA dust respectively. Black box in the insert shows the aperture where the 1D spectrum (Fig. \ref{fig:spec_full}) is extracted. }
    \label{fig:IFU+spec}
\end{figure*}

\begin{figure*}
    \centering
    \includegraphics[width=\textwidth,clip]{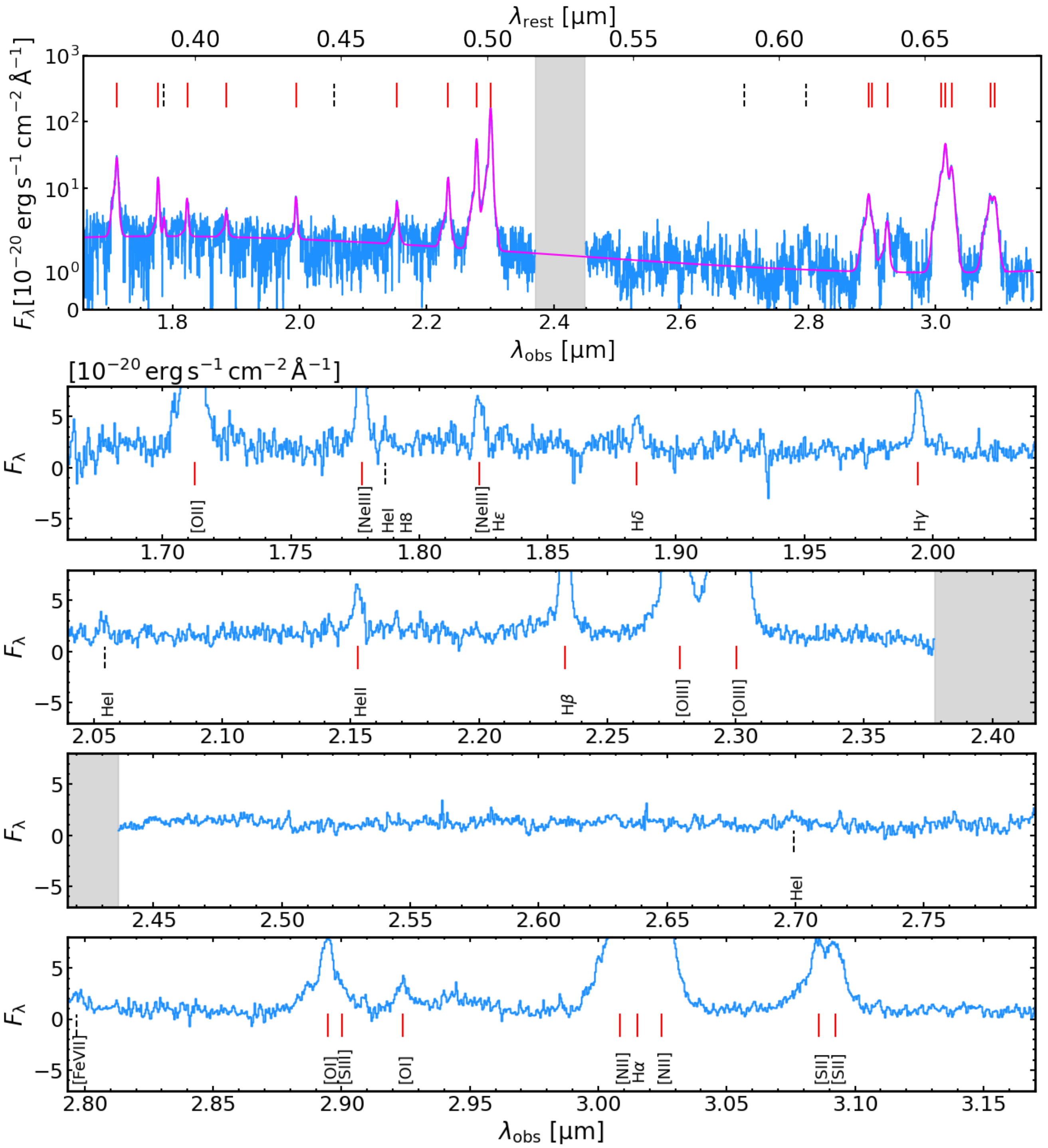}
    \caption{Full spectrum extracted at the AGN position without continuum subtraction (black box in Fig. \ref{fig:IFU+spec}d). We mark the emission lines detected with S/N$\gtrsim10$ in red lines. Lines detected with $3<$S/N$<10$ and visually checked to be extended are marked with grey lines. We note that the feature at $\sim2.945\,\mu$m is likely instrumental due to the leakage flux from the MSA. The central grey shaded region is the detector gap. The top panel is the full view of the spectrum while the zoom-in view of the faint lines are shown in the bottom panels. The magenta curves show the best fit of the spectrum (Appendix \ref{app:linelist}).}
    \label{fig:spec_full}
\end{figure*}

\section{4C+19.71, observation and data processing} \label{sec:data_process}

\object{4C+19.71} (\object{MG 2144+1928}) is the first target observed as part of the \textit{JWST} Cycle 1  General Observers program (ID 1970, PI: Wuji Wang) ``Zooming into the monster's mouth: tracing feedback from their hosts to circumgalactic medium in $z=3.5$ radio-loud AGN". The program includes four HzRGs with a diversity of star formation rates and radio jet morphologies \citep[][]{vanOjik_1996,Carilli_1997,Nesvadba_2007a,Falkendal_2019}. We selected them to have similar redshifts ($z\simeq3.5$) to maximize the number of emission lines in a single observational setting. 

4C+19.71 has a FR II-type radio jet with a double-sided lobe extending over $\gtrsim 60$ kpc \citep[][]{Fanaroff_1974,Carilli_1997}. Besides the radio data, there are multi-wavelength studies available for 4C+19.71 \citep[e.g.,][]{Pentericci_1999,Maxfield_2002,Seymour_2007,DeBreuck_2010}. The HzRG is surrounded by an X-ray halo and [\ion{O}{iii}]$5007$ nebula with similar extension. Both are elongated in the same direction as the jet \citep[$\sim 60$ kpc;][]{Armus_1998,Smail_2012,Nesvadba_2017a}. A more extended, $\sim 100$ kpc Ly$\alpha$ nebula is also detected around the radio AGN. It also shows an elongated morphology along the jet axis, even beyond the radio lobes \citep[][]{Falkendal_2021,wang2023}.

The stellar mass of the host galaxy of 4C+19.71 is constrained to be $\lesssim 10^{11.1}\,\mathrm{M_{\sun}}$. It forms stars at the rate of $\sim84\,\mathrm{M_{\sun}\,yr^{-1}}$ \citep[][]{Falkendal_2019}. Finally, there is significant molecular gas in the host \citep[$2.5\times 10^{10}\,\mathrm{M_{\sun}}$,][]{kolwa2023}.

The observation of 4C+19.71 was conducted on UT 2022 October 30 with NIRSpec in IFU mode \citep[][]{Boeker_2022,Jakobsen_2022}. We selected G235H/F170LP as the disperser/filter setup\textit{,} which covers $1.70-3.15\,\mathrm{\mu m}$ in the observed frame or $\sim3700-7620\,\mathrm{\AA}$ in the rest frame. The combination of filter and grating offers the spectral resolution of $85-150\,\mathrm{km\,s^{-1}}$. This ensures major optical emission lines seen in type-2 AGN ([\ion{O}{ii}]$3726,\,3729$ to H$\alpha$, including [\ion{S}{ii}]$6716,6731$) can be observed at once \citep[e.g.,][]{Zakamska_2003}. We adopted a 9-point dither pattern and the improved reference sampling and subtraction (IRS$^2$) read-out pattern. The total on-source exposure time was 3.7 h. We designed the leakage exposure, 0.4 h, to identify light leaking through the failed open shutters on the microshutter array (MSA) at the first dither position.

We downloaded the raw data from the Mikulski Archive for Space Telescopes (MAST). The data were processed with the {\em JWST} Science Calibration pipeline\footnote{\url{https://github.com/spacetelescope/jwst}} (v1.9.4) with the Calibration Reference Data System (CRDS) context file jwst\_1041.pmap. Our procedure is similar to \citet{Vayner_2023Q3D}, executing the standard first and second stages of the pipeline with one modification. Specifically, we used the ``emsm" method instead of ``drizzle" during the data cube construction step to reduce the low-frequency ripples due to undersampling. The third and final stage of the processing is to combine cubes from different exposures into the final cube. We instead used the script from \citet{Vayner_2023Q3D}, who applied the Python package \texttt{reproject} because the third stage of the pipeline falsely rejects the bright emission peak while keeping some noise spikes. During this step, sigma clipping ($2\sigma$) was also included to reject outliers. The data cube was resampled to have a pixel scale of 0.05\arcsec. We processed the raw NIRSpec IFU data of the standard star TYC 4433-1800-1 (PID 1128) for flux calibration. Finally, we performed an additional background subtraction to the flux-calibrated science data cube to alleviate negative background continuum seen in source-free regions. The size of the point spread function (PSF) of the NIRSpec IFU depends on wavelength and spatial position. We mark the full width at half maximum (FWHM) of the PSF constructed in \citet{Vayner_2023Q3D_2} in Fig. \ref{fig:IFU+spec}a.

Several works based on NIRSpec IFU observations report a World Coordinate System (WCS) offset in the final data cube \citep[e.g.,][]{Wylezalek_2022q3d,Perna_2023}. We aligned the continuum emission position of the west foreground galaxy in our NIRSpec cube to the Hubble Space Telescope (\textit{HST}) WFPC2 image and used this new WCS in the following analysis \citep[see Fig. \ref{fig:IFU+spec}a and][]{Pentericci_1999}. The astrometry of the \textit{HST} image was corrected by cross-matching with \textit{Gaia} DR3 targets \citep[][]{gaiaDR3_2023}. Our final WCS solution resulted in a shift of $\Delta$RA$=0.43\arcsec$ and $\Delta$Dec$=-0.22\arcsec$ compared to the pipeline output WCS. We discuss the astrometry correction in Appendix \ref{app:astrometry}. 

We also present the ALMA Band 8 observation of 4C+19.71. This observation was carried out under program ID 2021.1.00576.S (PI: Wuji Wang) on UT 2022 June 17. In this work, we use only the archived image at $\nu_{\rm obs}=$400.3 GHz to show the position of the cold dust emission. A detailed analysis of the ALMA observations is beyond the scope of this paper and will be presented in a forthcoming paper.

\begin{figure*}
    \centering
    \includegraphics[width=\textwidth,clip]{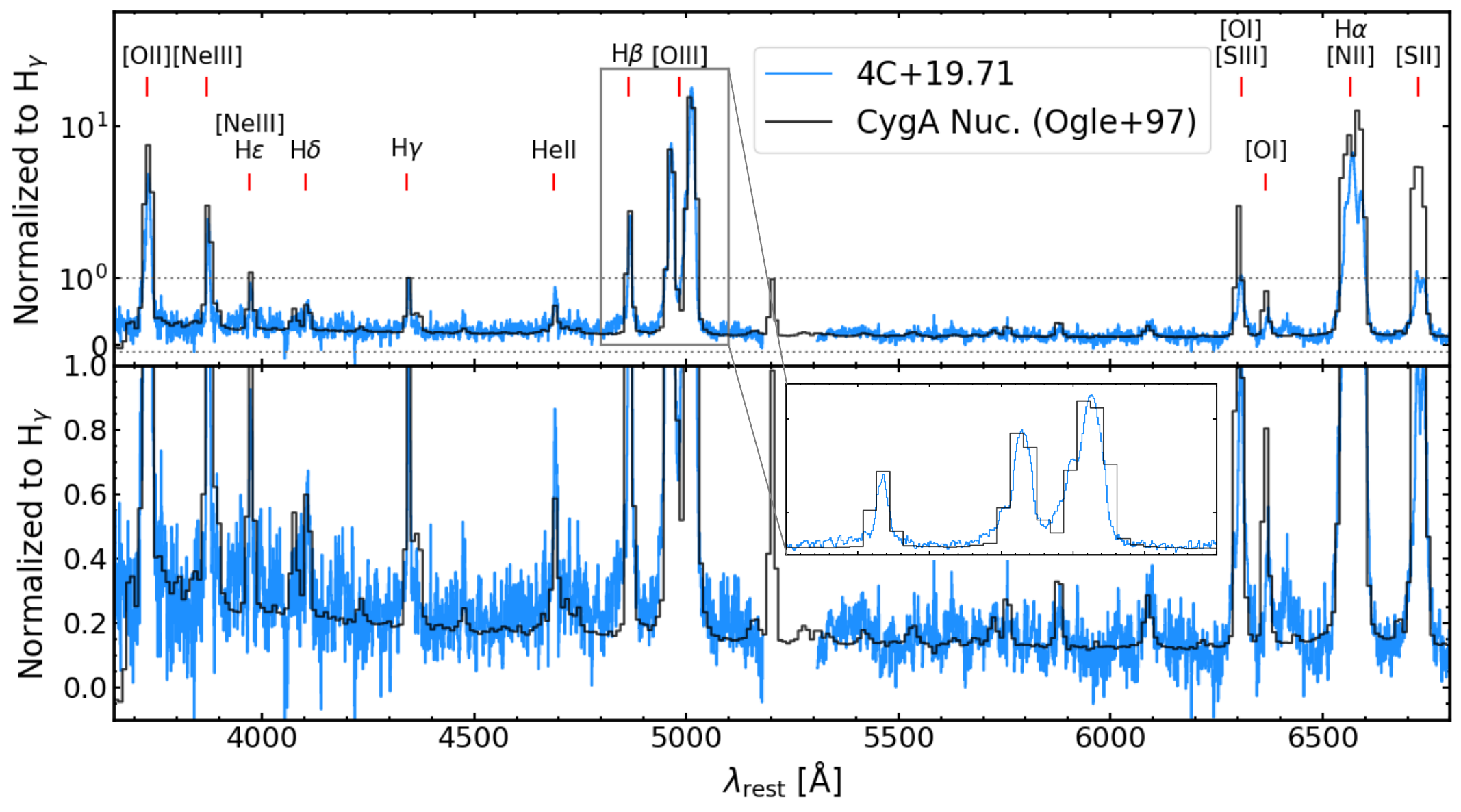}
    \caption{Spectral comparison of 4C+19.71 to Cygnus A \citep[][]{Ogle_1997}. Blue histogram shows the same 1D spectrum as in Fig. \ref{fig:spec_full}. The spectrum of Cygnus A is shown in black which is extracted at its nucleus position after Galactic reddening correction and subtraction of a elliptical galaxy template. Both spectra are normalised to the peak flux density of their H$\gamma$. The bottom panel is a zoom-in version of the region between the grey dotted lines in the upper panel. We marked the emission lines detected for 4C+19.71 with S/N$\gtrsim10$ as in Fig. \ref{fig:spec_full}. The inset provides a further zoom-in around H$\beta$-[\ion{O}{iii}] complex. }
    \label{fig:spec_comp}
\end{figure*}

\section{Results} \label{sec:results}
In this paper, we inspect both the 2D morphology and 1D spectrum of the warm ionized gas. Given the signal-to-noise ratio (S/N) of the optical emission lines covered in our setting, we choose the [\ion{O}{III}]5007 ($[\ion{O}{iii}]$ hereafter if not specified) as the proxy for the morphology. It is extensively used in the studies of AGN narrow line regions \citep[e.g.,][]{Husemann_2013b,Harrison_2014,Wylezalek_2016b}. This is the first time that the distribution of ionized gas around a $z>3.5$ radio loud AGN is seen  with $\lesssim1\,$kpc resolution. 

\subsection{Morphology of the extended ionized nebula}\label{subsec:O3morphology}

\citet{Armus_1998} observed the [\ion{O}{iii}] emission of 4C+19.71 with Keck narrow-band imaging. They reported the central [\ion{O}{iii}] (i.e., a similar region as observed in the field of view of the NIRSpec IFU) has an elongated morphology along the north-south direction with two distinctive parts. We show the first zoom-in view of the [\ion{O}{iii}] nebula of 4C+19.71 in Fig. \ref{fig:IFU+spec}. We produce three pseudo-narrow-band images and show their composite in Fig. \ref{fig:IFU+spec}a. The individual images are shown in Fig. \ref{fig:IFU+spec}(b1), (b2) and (b3) for the blue wing ([$-700,-500$]\,$\mathrm{km\,s^{-1}}$), systemic redshift ([$-100,100$]\,$\mathrm{km\,s^{-1}}$) and red wing ([$500,700$]\,$\mathrm{km\,s^{-1}}$), respectively. On each of the images, there are two distinct peaks located south and north which roughly align with the jet axis \citep[dark purple contours in Fig. \ref{fig:IFU+spec}a,][]{Carilli_1997}. Based on [\ion{O}{iii}] gas kinematics probed by VLT/SINFONI near the jet hot spot positions \citep[Fig. A.7 of][]{Nesvadba_2017b}, we assume the southern jet is the approaching one. There is no doubt that 4C+19.71 has a hidden quasar \citep[e.g.,][and Sect. \ref{subsec:low_z_comp}]{Seymour_2007,DeBreuck_2010,Falkendal_2019}. Assuming the ionization cone of the quasar is aligned with the jet axis \citep[][]{Drouart_2012}, it is likely that we are observing, in both the south and north regions, the foreground and background parts of the cone. 

To assist in studying the morphology, we show example [\ion{O}{iii}]$4959,5007$ spectra extracted at four different locations (Fig. \ref{fig:IFU+spec}(c1) to (c4)). We note that the analysis of the extended emission-line region kinematics is not the focus of this paper. The per-spatial-pixel (spaxel) spectra are fitted using \texttt{q3dfit} \citep[][]{Rupke_2023q3dfit}, which is a Python tool for analyzing \textit{JWST} IFU data based on the software \texttt{IFSFIT} \citep[][]{Rupke_2014,Rupke_2021}\footnote{\url{https://q3dfit.readthedocs.io/en/latest/}}. It is clear that at least three different kinematic components are present on galactic scales ($\lesssim$10$\,$kpc). This is especially obvious for the gas in the south, with three distinct peaks present (e.g., Fig. \ref{fig:IFU+spec} (c2)). We stress that we refer to the component closest to $0\,\mathrm{km\,s^{-1}}$ as ``systemic'' and the others as ``blue'' or ``red'' depending on their relative velocity shift with respect to the systemic component. At this point, it remains unclear whether they are the blueshifted or redshifted outflowing gas clouds.

The most striking morphological features are the spatially extended emission blobs and filamentary structures around the systemic velocity (Fig. \ref{fig:IFU+spec}(b1)). For example, both components detected in the east filament (Fig. \ref{fig:IFU+spec}(c1)) and the gas of the northern spot (Fig. \ref{fig:IFU+spec}(c3)) all have $|\Delta v| <300\,\mathrm{km\,s^{-1}}$. Neither the previous narrow-band imaging nor ground-based IFU could resolve these components spatially and spectrally \citep[][]{Armus_1998,Nesvadba_2017b}. With the relatively small field of view (FoV) of the NIRSpec IFU, the jet hot spots are not captured which are presumed to be the regions with the most energetic jet-gas interactions. Given the fact that these extended filamentary features sit at the systemic redshift, we arrive at the conclusion that they are not associated with the outflow but may be high density gas clumps. Their morphologies may have been disrupted by the jet-induced bubble \citep[][]{Mukherjee_2016,Dugan_2017}. 
Indeed, at least in the example spectra extracted at the east filament, we observe a broad component indicative of disruption by the jet (FWHM$ = 1052\,\mathrm{km\,s^{-1}}$).

\begin{figure*}
    \centering
    \includegraphics[width=\textwidth,clip]{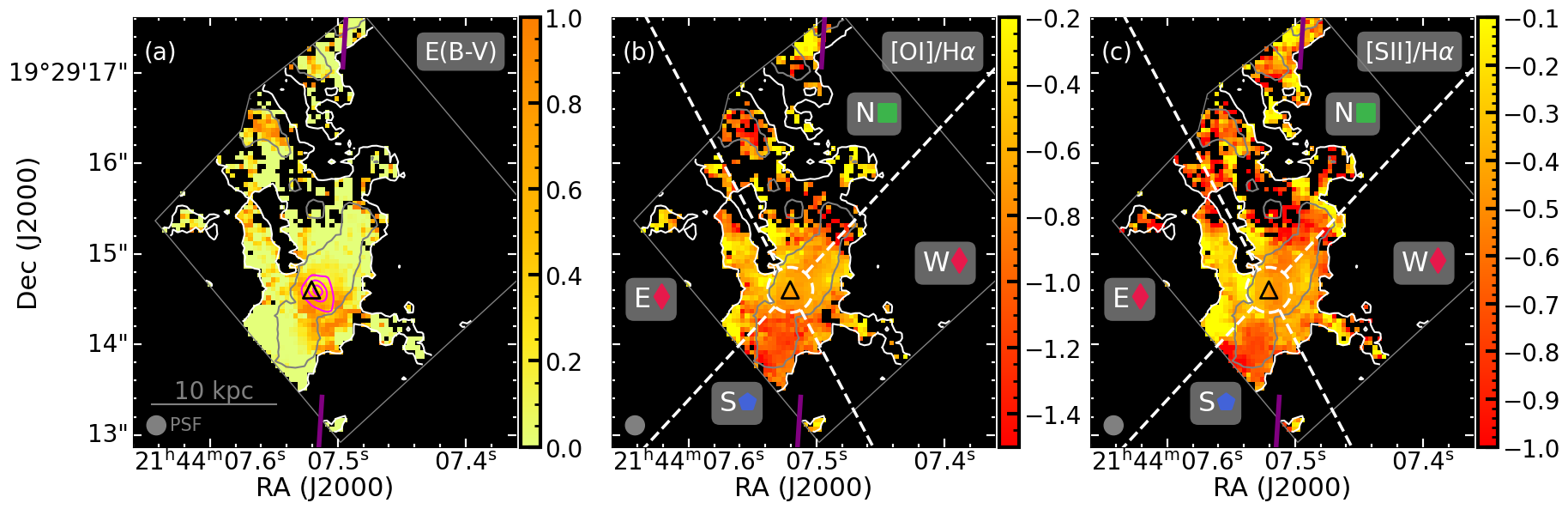}
    \caption{Dust attenuation (a) and line ratio maps (b, c) constructed based on the fitted fluxes integrated for all components. The white and grey contours indicate the $5\sigma$ and $15\sigma$ [\ion{O}{iii}] surface brightness levels of the un-smoothed systemic narrow-band image, respectively (Fig. \ref{fig:IFU+spec}). Dark purple lines show the directions toward the radio jet hot spots. Black triangle marks the position of the AGN while magenta contours trace the ALMA continuum (Fig. \ref{fig:IFU+spec}d). The dashed white lines and circle in panel (b) and (c) divide the FoV into five regions for line ratio diagnostic analysis (Fig. \ref{fig:q3dfit_line_ratio}). FWHM of the \textit{JWST} PSF is marked at the left corner.}
    \label{fig:q3dfit_line_ratio_map}
\end{figure*}

\begin{figure*}
    \centering
    \includegraphics[width=\textwidth,clip]{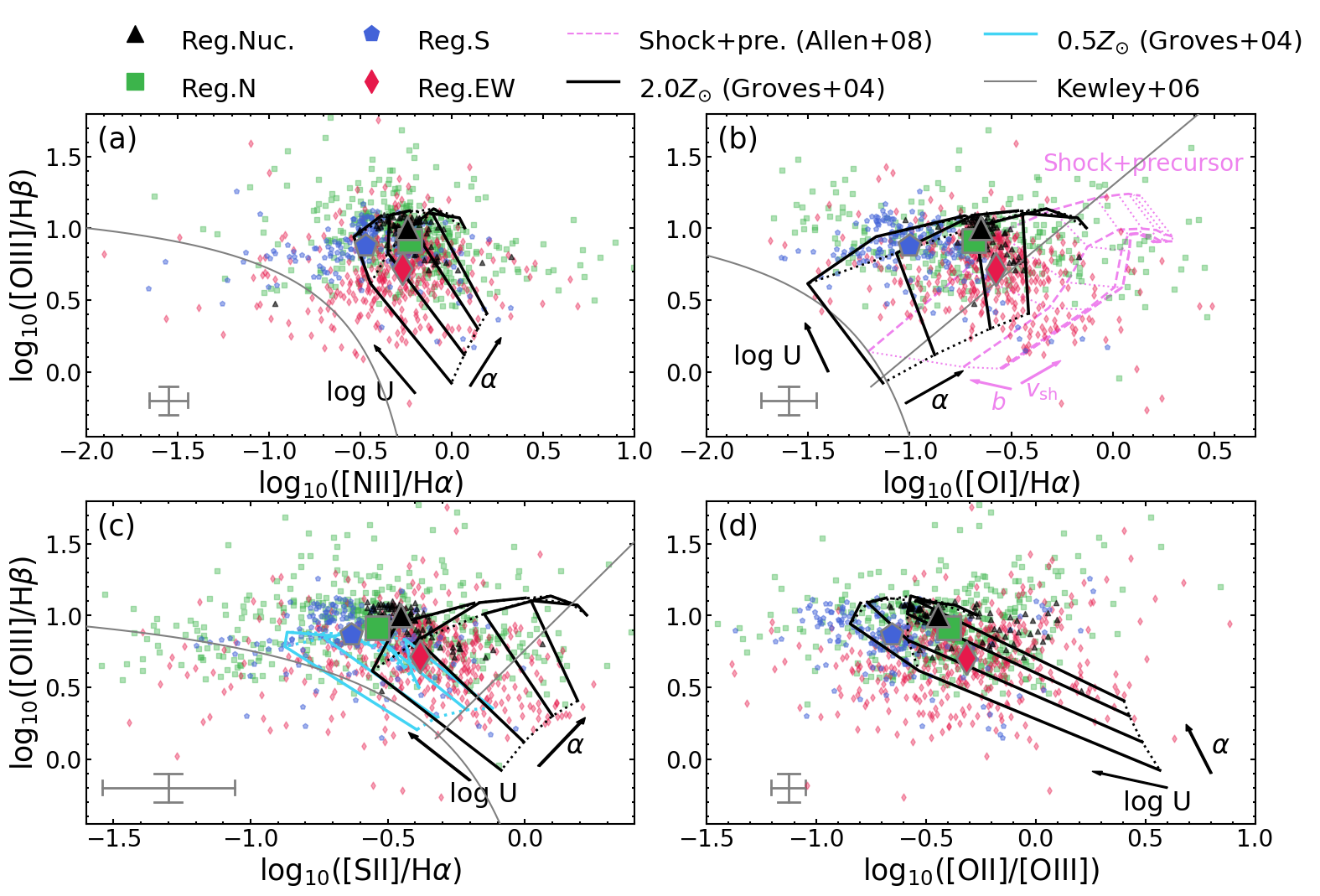}
    \caption{Line ratio diagnostic diagrams based on fitting and corrected for dust attenuation. Each smaller mark corresponds to one spaxel in the FoV. The larger symbols represent the median. Data points with the same symbol are the same region marked as Fig. \ref{fig:q3dfit_line_ratio_map}: Nuc. --- nucleus (circle with $r=0.25\arcsec$), N --- north, S --- south, EW --- east+west. The typical $1\sigma$ error is shown at the left corner. We show the $2Z_{\odot}$ dusty radiation-pressure dominated photonionization models from \citet{Groves_2004_model} in black and mark the direction of increasing ionization parameter, $\log U \in [-3.0, -2.0, -1.0, 0.0]$, and power-law index, $\alpha \in [-2.0, -1.7, -1.4, -1.2]$. The similar model with $0.5Z_{\odot}$ is shown in panel (c) in cyan. Purple grids are the ``shock+precursor'' models with per-shock density $100\,\mathrm{cm^{-3}}$, \citep[shock velocities of $200\leq v_{\rm shock}\leq1000\,\mathrm{km\,s^{-1}}$ in steps of 100$\,\mathrm{km\,s^{-1}}$ and magnetic parameters of $0.01\leq b=B/n^{1/2} \leq 100\,\mathrm{\mu G\,cm^{-3/2}}$ in steps of 1 dex, ][]{Allen_2008}. The directions of increasing $v_{\rm shock}$ and $b$ are indicated with purple arrows. We include the empirical classification from \citet{Kewley_2006} in grey lines in panels (a), (b) and (c).}
    \label{fig:q3dfit_line_ratio}
\end{figure*}
 
\subsection{Continuum emission}\label{subsec:wcsresults}

We overlay in Fig. \ref{fig:IFU+spec}b the continuum emission detected in our ALMA Band 8 observation, which likely indicates the location of the cold dust emission heated by the newly formed stellar population \citep[e.g.,][]{Herrera-Camus_2021}. In Fig. \ref{fig:IFU+spec}d we show that the AGN position (NIRSpec continuum, Sect. \ref{subsec:low_z_comp}) agrees with the ALMA dust location within their uncertainties. The $1\sigma$ uncertainty, $\sim0.02\arcsec$, of the ALMA dust position is estimated using $\mathrm{pos_{acc}=beam_{FWHP}/(S/N)/0.9}$ with $\rm beam_{FWHP}\simeq0.2\arcsec$ and S/N$\simeq10$ (see Appendix \ref{app:astrometry} for the $1\sigma$ uncertainty, $\sim0.06\arcsec$, of the NIRSpec continuum) \footnote{ \url{https://statics.teams.cdn.office.net/evergreen-assets/safelinks/1/atp-safelinks.html}}.

A previous \textit{HST} image (WFPC2/F702W) indicates other two objects in the NIRSpec IFU FoV \citep[][]{Pentericci_1999}. They are also seen in our NIRSpec observations (Fig. \ref{fig:IFU+spec}a). We label them G1 and G2 for the galaxy in the west and east, respectively. G1 is used as the WCS alignment reference in Section \ref{sec:data_process}. They are identified as foreground galaxies with $z\simeq1.786$ for G1 and $z\simeq1.643$ for G2. Their spectra and emission line identifications are shown in Appendix \ref{app:fore_galaxy}.

\subsection{Emission line identification} \label{subsec:line_emission}

\begin{table}[]
    \centering
        \caption{Fitted kinematic results from the 1D spectrum at nucleus.  }
    \label{tab:1Demission_kine}
    \begin{tabular}{lccc}

    \hline
    \hline
     & central comp. & red comp. &  blue comp. \\
   
    \hline
    $\Delta v\, [\mathrm{km\,s^{-1}}$] & 238$\pm10$ & 572$\pm30$ & -148$\pm20$\\
    \hline
     FWHM$\, [\mathrm{km\,s^{-1}}$]   & 445$\pm10$ &  412$\pm30$ & 1669$\pm20$ \\
    \hline

    \hline
    \end{tabular}
    \tablefoot{The velocity shift is with respect to the systemic redshift from \citet{kolwa2023}, $z=3.5892$. }

\end{table}

We present the identification of the emission lines in Fig. \ref{fig:spec_full}. This is the first time that optical nebular emission lines from [\ion{O}{ii}]$3726,3729$ to [\ion{S}{ii}]$6716,6731$ are observed for $z\simeq3.5$ HzRGs with the spectral resolution of $85$ to $150\,\mathrm{km\,s^{-1}}$. This makes it possible to investigate the ionization mechanisms and metallicities \citep[e.g., line ratio diagnostic diagrams,][]{BPT_1981,Veilleux_1987,Kauffmann_2003,Groves_2004,Groves_2006,Kewley_2001,Kewley_2006,Kewley_2013,Wylezalek_2017}. 

We extracted a spectrum from a $0.2\arcsec\times0.2\arcsec$ or $1.4\times1.4\,\mathrm{kpc}^{2}$ ($4\times4$ pix$^{2}$) square aperture near the position of the continuum emission peak. The aperture and size were chosen to also include the [\ion{O}{iii}] emission peak which maximizes the S/N for emission line identification. The aperture was aligned with the extension of the brightest [\ion{O}{iii}] emission blob. We show this aperture in the black box in Fig. \ref{fig:IFU+spec}d. The spectrum is presented in Fig. \ref{fig:spec_full}. We summed over five wavelength pixels around the systemic zero (velocity range of $\sim270$ to $\sim150\,\mathrm{km\,s^{-1}}$) to calculate the line $S/N$ for identification. We identify $\sim24$ emission lines (doublets) with S/N$>3$. The ones with S/N$\gtrsim10$ are marked by red lines. We implemented a visual check by examining the spatial distribution of lines with $3<$S/N$<10$. The spatially resolved ones are marked by black dashed lines.

We fitted the spectrum with \texttt{q3dfit}. We used three kinematic components. The line center and width of the same component are connected for different lines. We present these kinematic results in Table \ref{tab:1Demission_kine}. We refer to them as central, red and blue components based on their velocity shifts. In Appendix \ref{app:linelist}, we report fitted line fluxes from the 1D spectrum.

\section{Discussion and Summary} \label{sec:summary}

\subsection{Comparison to Cygnus A}\label{subsec:low_z_comp}
Cygnus A (3C405) is the most powerful radio AGN in the local Universe with comparably high radio power as 4C+19.71 \citep[$z=0.0562$, $P_{\rm 178\,MHz}\sim 5.5\times10^{28}\,\mathrm{W\,Hz^{-1}}$,][]{Carilli_1996}. This makes it the perfect target to compare with the properties of the warm ionized ISM gas using the common optical tracers \citep[][]{Fosbury_1999,Vernet_2001_thesis}. Using the spectropolarimeter on the Keck telescope, \citet{Ogle_1997} observed the nebular emission lines of Cygnus A. We overlay our 1D spectrum (Fig. \ref{fig:spec_full}) with the spectrum of Cygnus A extracted at its nucleus in Fig. \ref{fig:spec_comp}. The extraction aperture of Cygnus A is $\sim1"\times1.1"$ which corresponds to $1.1\times1.2\,\mathrm{kpc}^{2}$ and is similar to the aperture used for 4C+19.71 ($1.4\times1.4\,\mathrm{kpc}^{2}$). This is the first time that a rest-frame optical spectrum of a $z\gtrsim 1$ radio AGN can be studied in comparable detail as a local example. 

To first order, the continuum shape of 4C+19.71 is consistent with Cygnus A. We note that the Cygnus A spectrum is corrected for Galactic reddening and had its host galaxy continuum subtracted \citep[][]{Ogle_1997}. Though there might be contamination of continuum  (i.e., due to the leak of from the MSA) for our NIRSpec data, we conclude that the emission from evolved stellar population in the host of 4C+19.71 is faint, i.e., the spectrum is dominated by the (possibly scattered) light of the hidden AGN. This indicates that the stellar population has a sub-dominant contribution in the regions near the AGN.

The relative emission line fluxes are comparable for 4C+19.71 and Cygnus A at least from [\ion{O}{ii}]$3726,3729$ to [\ion{O}{iii}]. We conclude that, to first order, AGN photoionization mechanisms dominate the inner parts of these two galaxies. The similarity of the Balmer lines, H$\beta$, H$\gamma$ and H$\delta$ suggests the ISM gas conditions are similar \citep[][]{Osterbrock_2006}. The dissimilarity of the flux at redder wavelength (e.g., [$\ion{O}{i}$], [$\ion{N}{ii}$] and [$\ion{S}{ii}$]) may suggest differences in gas enrichment. 
\citet{vanBemmel_2003} proposed that the quasar wind-driven outflowing dust clouds scatter the optical line emitting photons in Cygnus A. We know there are at least three kinematic components present in the nucleus of 4C+19.71 (Sect. \ref{subsec:O3morphology}). We defer to a future publication the detailed study of the gas kinematics to unveil the spatially resolved gas motions and constrain the ISM distribution.

\subsection{Line ratio diagnostics and ionization mechanisms}\label{subsec:linera_diag}

To study the extended irregular ISM structures around the hidden powerful radio AGN (Sect. \ref{subsec:O3morphology}), we construct line ratio maps and compare them to emission line diagnostics \citep[e.g.,][]{Groves_2004,Kewley_2006,Groves_2006}.  We use \texttt{q3dfit} to fit the cube to systematically separate the blended emission lines and account for doublets. The fitted region was selected based on the [\ion{O}{iii}] narrow-band image with a S/N cut of $>5$ (white contour in Fig. \ref{fig:q3dfit_line_ratio_map}). We limit the spatial study in this paper to the six brightest emission lines: [\ion{O}{ii}]$3726,3729$, H$\beta$, [\ion{O}{iii}]$4959,5007$, [\ion{O}{i}]$6300$, [\ion{N}{ii}]$6548,6583$, H$\alpha$ and [\ion{S}{ii}]$6716,6731$. These lines are fitted simultaneously with the same velocity shift and line width for the corresponding components \citep[``kinematically tied", e.g.,][]{Zakamska_2016b}. Following \citet{Vayner_2023Q3D}, we ran the fit with 3, 2 and 1 as the maximum number allowed Gaussian components. We performed a visual inspection to determine the best fit for each spaxel. In the following discussions, we sum the the fluxes of the doublet [\ion{O}{ii}]$3726,3729$ and [\ion{S}{ii}]$6716,6731$, respectively. We use the fluxes of the individual [\ion{O}{iii}]$5007$ and [\ion{N}{ii}]$6583$ lines, respectively. Hereinafter, we omit the wavelength when referring to the line name. The line ratio maps and diagnostics based on the fit are shown in Fig. \ref{fig:q3dfit_line_ratio_map} and \ref{fig:q3dfit_line_ratio}, respectively. As mentioned before, a kinematical study of the ionized ISM is beyond the scope of this paper; we only derive the line ratios of the integrated line fluxes of all velocity components.

The Balmer line ratios are sensitive to dust attenuation \citep[e.g.,][]{Veilleux_1987,Osterbrock_2006}. We present the color excess, $E(B-V)$, maps based on the fitted H$\alpha$/H$\beta$ flux ratios in Fig. \ref{fig:q3dfit_line_ratio_map}a. The map is constructed using an intrinsic flux ratio of 3.1 assuming Case B recombination in AGN ionization and a \citet{Calzetti_2000} extinction law following \citet{Vayner_2023Q3D}.  We find that there is a dusty region with high $E(B-V)$$\sim1$ around the AGN position, coincident with our ALMA dust continuum (pink contours in Fig. \ref{fig:q3dfit_line_ratio_map}a), confirming the importance of dust causing extinction near the AGN. This spatial coincidence also validates our manual astrometry correction based on foreground galaxy (Appendix \ref{app:astrometry}). \citet{Fosbury_1999,Vernet_2001_thesis} proposed that a dust ring is present around the nucleus of Cygnus A. Given the spectral similarity of 4C+19.71 to Cygnus A (Fig. \ref{fig:spec_comp}), it is possible that we also detect a dust lane around the center of 4C+19.71. The line ratios shown in Fig. \ref{fig:q3dfit_line_ratio_map}bc are corrected for the dust attenuation based on this $E(B-V)$ map.

In Fig. \ref{fig:q3dfit_line_ratio_map}b and c, we show respectively the ratio maps of $\mathrm{[\ion{O}{i}]/H\alpha}$ and $\mathrm{[\ion{S}{ii}]/H\alpha}$ based on the integrated line fluxes of all components. The following line ratios are in logarithmic scale, unless otherwise specified. We group the spaxels in the FoV into five regions (white dashed lines in Fig. \ref{fig:q3dfit_line_ratio_map}bc), following the apparent geometry of the AGN ionization cone, with the nucleus added as an individual region. The southern part of the ISM is distinctive as having lower $\mathrm{[\ion{O}{i}]/H\alpha}$ and $\mathrm{[\ion{S}{ii}]/H\alpha}$ values.  In the extended filamentary part of the nebula toward the north, there is a large scatter with the presence of both higher and lower $\mathrm{[\ion{O}{i}]/H\alpha}$ and $\mathrm{[\ion{S}{ii}]/H\alpha}$ values. In general, the east and west regions have higher $\mathrm{[\ion{O}{i}]/H\alpha}$ and $\mathrm{[\ion{S}{ii}]/H\alpha}$. To simplify, we combine the East and West regions and refer to it as EW.

We inspect our results in line ratio diagnostic diagrams and present in Fig. \ref{fig:q3dfit_line_ratio}. In addition to $\mathrm{[\ion{O}{i}]/H\alpha}$ (Fig. \ref{fig:q3dfit_line_ratio}b) and $\mathrm{[\ion{S}{ii}]/H\alpha}$ (Fig. \ref{fig:q3dfit_line_ratio}c), we present the $\mathrm{[\ion{N}{ii}]/H\alpha-[\ion{O}{iii}]/H\beta}$ (Fig. \ref{fig:q3dfit_line_ratio}a) and $\mathrm{[\ion{O}{ii}]/[\ion{O}{iii}]-[\ion{O}{iii}]/H\beta}$ diagnostics (Fig. \ref{fig:q3dfit_line_ratio}d). The east$+$west (EW) region (red diamonds) occupies different locations on the diagnostic diagrams than the other regions, with higher $\mathrm{[\ion{O}{i}]/H\alpha}$, $\mathrm{[\ion{S}{ii}]/H\alpha}$ and $\mathrm{[\ion{O}{ii}]/[\ion{O}{iii}]}$ values. The EW region is also distinctive with a lower $\mathrm{[\ion{O}{iii}]/H\beta}$ compared to the other regions. Although the southern spaxels (blue hexagons) are clustered toward lower $\mathrm{[\ion{N}{ii}]/H\alpha}$, $\mathrm{[\ion{O}{i}]/H\alpha}$, $\mathrm{[\ion{S}{ii}]/H\alpha}$ and $\mathrm{[\ion{O}{ii}]/[\ion{O}{iii}]}$ values, we find they fall in the line ratio space spanned by the northern regions (green squares) in all diagrams. Despite having a relatively large dispersion, the median line ratios of the northern region is consistent with the spaxels from the nucleus region (black triangles). We report the median line ratios of each region in Table \ref{tab:Line_diag_median}.


We overlay the \citet{Groves_2004_model} dusty radiation-pressure dominated model grids on the diagnostic diagrams. We chose the $2Z_{\odot}$ models (black) as indicated by the $[\ion{N}{ii}]/\mathrm{H\alpha}$ ratios \citep[e.g., metallicity from][]{Groves_2006,Nesvadba_2017b}. The models are constructed with various combinations of the power-law index, $\alpha$, of the ionizing source (i.e., $F_{\nu}\propto\nu^{\alpha}$) and ionization parameter, $U=S_{\star}/(n_{\rm H}c)$, where $c$ is the speed of light, $n_{\rm H}$ is the hydrogen number density and $S_{\star}$ is the flux of ionizing photons entering the cloud. The goal is not to perform a quantitative comparison, but rather to test which parameter(s) could be responsible for the line ratio differences. Although other interpretations are possible, a higher $U$ parameter could explain the lower $[\ion{O}{i}]/\mathrm{H\alpha}$ and a higher $\mathrm{[\ion{O}{iii}]/H\beta}$ \citep[e.g.,][]{villar_martin1999}. Our comparison indicates a scenario where the north, south and nucleus regions have higher $U$ parameters than the EW region. This is expected given that the jet axis is along the north-south direction which is also the direction of the AGN ionization cone \citep[][]{Drouart_2012}. Hence, the ionizing photons of the AGN can reach the ISM with less extinction along the North-South direction. In this configuration, the East and West regions are outside of the ionization cone where the photons will encounter more obstacles before ionizing the gas \citep[e.g.,][]{Fosbury_1999,vernet_2001}. This indicates larger particle density $n_{\rm H}$ and less ionizing photon flux (smaller $S_{\star}$) in the EW which leads to lower observed $U$ parameters. 

The south region has lower $\mathrm{[\ion{N}{ii}]/H\alpha}$, $\mathrm{[\ion{O}{i}]/H\alpha}$, $\mathrm{[\ion{S}{ii}]/H\alpha}$ and $\mathrm{[\ion{O}{ii}]/[\ion{O}{iii}]}$. The $U$ parameter appears at similar levels in the south as in the nucleus. Comparing with the model grids, this could indicate that the ratio difference is due to lower $\alpha$ (steeper power-law ionizing spectrum). If this is the case, then there may be additional extinction between the ionizing photons and the southern clouds. The attenuation map (Fig. \ref{fig:q3dfit_line_ratio_map}a) indeed shows a higher value toward the south. If this is due to the geometry of the dust torus, it may be inconsistent with the indicated orientation from \citet{Nesvadba_2017b}. We also overlay the similar radiation model with $0.5Z_{\odot}$ in cyan in Fig. \ref{fig:q3dfit_line_ratio}c and find that there is an alternative possibility that the southern region is more metal poor than the nuclear region. Additionally, the complex kinematics in the southern region (Fig. \ref{fig:IFU+spec} (c3)), suggesting a more complex scenario in the south (e.g., a late-stage merger, see Sect. \ref{subsec:southISM}). The detailed analysis of this southern region requires more lines which is beyond the scope of this paper \citep[e.g., He lines,][]{Groves_2004_model,Holden_2023a}. 

To further test the ionization mechanisms, we also overlay the empirical classification from \citet{Kewley_2006} in Fig. \ref{fig:q3dfit_line_ratio}a, b and c. Although these are calibrated using low-$z$ data \citep[i.e., may not be relaible for high-$z$, see][]{Kewley_2019}, the results provide evidence of the AGN radiation dominating ionization of the ISM within $\lesssim30\,$kpc of the hidden quasar. Some spaxels in the north and east regions are located in the low ionization or \ion{H}{ii} region. This suggests a different ionization mechanism or mix of different mechanisms at these spatial locations. Indeed, we find a part of them are roughly consistent with the ``shock+precursor'' models \citep[light purple grids Fig. \ref{fig:q3dfit_line_ratio}b,][]{Allen_2008}. The shock models can cover a large fraction of the parameter space of the line ratio diagnostics. The identification of shock ionization requires combination of flux ratios with kinematics, electron temperature and H$\alpha$ fluxes \citep[e.g.,][]{Alatalo_2016,Baron_2017,Alarie_2019}. Our discussion is not meant to be decisive but to be inclusive with the possibility of shock. We choose the models with pre-shock number density, $n=100\,\mathrm{cm^{-3}}$ \citep[e.g.,][]{DeBreuck_2000}. For the shock velocities, we use higher values ($v_{\rm shock}\geq200\,\mathrm{km\,s^{-1}}$) following \citet{DeBreuck_2000} studying the emission lines of HzRGs. We note that shock+precursor model with pre-shock density  $n=1000\,\mathrm{cm^{-3}}$ are also physical for AGN \citep{Liu_weizhe_2020} which occupies a subset of the line ratio space spanned by $n=100\,\mathrm{cm^{-3}}$ model using the same set of $v_{\rm shock}$ and magnetic parameters. The driver of the shock is unclear. The EW region is roughly in the direction perpendicular to the ionization cone (jet axis) which suggests a less collimated shock driver or the ionization cone is not perpendicular to the dusty obscuring disk \citep[e.g.,][]{Dugan_2017,Tanner_2022,Vayner_2023Q3D}. As for the more extended parts ($\gtrsim10\,$kpc) in the North, the passage of the powerful radio jet could also have left remaining shock effects.

Our results show that the inner part of the ionized ISM in the $z\sim3.5$ radio-loud AGN is dominated by photons from the hidden AGN with potential evidence for shock ionization. Its structure agrees with the classical ionization cone. Although the analysis is based on knowledge gained at low-$z$ and there are more unsolved complexities, \textit{JWST} will be the key to address this \citep[e.g.,][]{Nakajima_2022,Sanders_2023}.

\begin{table}[]
    \centering
        \caption{Median line ratio values for different regions in Fig. \ref{fig:q3dfit_line_ratio}.}
    \label{tab:Line_diag_median}
    \begin{tabular}{lcccc}

    \hline
    \hline
      & Nuc. & N &  S & EW \\
    \hline
    $\mathrm{[\ion{O}{iii}]/H\beta}$ & 1.00 & 0.92 & 0.88 & 0.72 \\
    $\mathrm{[\ion{N}{ii}]/H\alpha}$ & -0.24& -0.23& -0.47& -0.27 \\
    $\mathrm{[\ion{O}{i}]/H\alpha}$  & -0.65& -0.68& -1.00& -0.58 \\
    $\mathrm{[\ion{S}{ii}]/H\alpha}$ & -0.45& -0.54& -0.63& -0.38 \\
    $\mathrm{[\ion{O}{ii}]/[\ion{O}{iii}]}$ & -0.44& -0.40& -0.65& -0.31 \\
    \hline

    \hline
    \end{tabular}

\end{table}

\subsection{Inefficient quasar-driven outflow at nucleus}\label{subsec:ouflow}

Our NIRSpec IFU observations enable studying outflows on scales $\lesssim20\,$kpc. VLT/SINFONI observations captured more extended outflowing  gas \citep[][]{Collet_2016,Nesvadba_2017b,Nesvadba_2017a}.


In Sect. \ref{subsec:O3morphology}, we show that the ``high-velocity" clouds are only detected within the $\sim5\,$kpc from the AGN and their morphology is aligned with the jet axis (Fig. \ref{fig:IFU+spec}). This is inconsistent with the results from observations of quasars at Cosmic Noon where fast ($|\Delta v|\sim500\,\mathrm{km\,s^{-1}}$) outflows are seen at $\gtrsim10\,$kpc and are less directionally confined \citep[e.g.,][]{Vayner_2023Q3D_2}. However, \citet{Cresci_2023,Veilleux_2023} also showed one contradicting case. Hence, this discussion is yet settled and required more observations. Through our discussions in Sect. \ref{subsec:low_z_comp} and \ref{subsec:linera_diag}, these may be the outflows inside the ionization cone.  The double-sided radio lobes have a projected extent of $\sim60\,$kpc, which is beyond the \textit{JWST} FoV.  Hence, this potential outflow maybe radiatively-driven. 


The systematic per-spaxel kinematic study is beyond the scope of this work. Additionally, as shown in Sect. \ref{subsec:linera_diag} (also in Sect. \ref{subsec:southISM}), there may be more complex scenarios happening in the southern cone. Hence, we use the results from the integrated 1D spectrum (Fig. \ref{fig:IFU+spec}d and \ref{fig:spec_full} and Table \ref{tab:1Demission_kine} and \ref{tab:emissionline}) at the nucleus region as a proxy to estimate the kinetic energy of the potential outflow near the AGN. For this calculation, we focus on the red and blue components (Table \ref{tab:1Demission_kine} and \ref{tab:emissionline}) and assume them to be the redshifted and blueshifted outflow clouds seen on the back and front side of the northern part of the ionization cone, respectively. Then, we assume their relatively velocity shifts with respect to the central component (Appendix \ref{app:sys_z}) as their outflow velocities ($\Delta v_{\rm red}=334\,\mathrm{km\,s^{-1}}$ and  $\Delta v_{\rm blue}=-386\,\mathrm{km\,s^{-1}}$). We follow \citet{Vayner_2023Q3D_2} and use the following equation to calculate the ionized outflow gas mass \citep[][]{Osterbrock_2006}:
\begin{equation}\label{eq:m_ion}
    M_{\text {ion}}=1.4\left(\frac{m_p L_{\rm H \alpha}}{j_{H_\alpha} n_e}\right),
\end{equation}
where $L_{\rm H\alpha}$ and $n_{e}$ are the H$\alpha$ luminosity and electron density, respectively. We assume an electron temperature $T_{e}\simeq15\,000\,$K \citep[][]{Nesvadba_2017b,Vayner_2023Q3D_2}. We use the fiducial value $n_{e}=500\,\mathrm{cm^{-3}}$ from \citet{Nesvadba_2017b}. Given that the emissivity, $j_{H_\alpha}$, is not sensitive to $n_{e}$, we use the value of $2\times 10^{-25} \mathrm{erg\, cm^3\,s^{-1}}$ for $T_{e}\simeq15\,000\,$K \citep[][]{Vayner_2023Q3D_2}. Based on equation \ref{eq:m_ion}, we find $M_{\rm ion} = 10^5$ and $10^6$~M$_\odot$\, for the red and blue components, respectively.
We then estimate the outflow rate using the following equation \citep[][]{Vayner_2023Q3D_2}:
\begin{equation}
    \dot{M}_{\rm ion}=\frac{M_{\rm ion} v_{\rm ion}}{\Delta R}.
\end{equation}
For the distance $\Delta R$, we use half of the physical size of the extraction aperture, $0.7\,$kpc. For the velocity, we use the $v_{\rm ion} = |\Delta v|+\sigma_{v}$ \citep[][]{Rupke_2013,Vayner_2023Q3D_2}. We report the summation of the red and blue components as the estimated total outflow rate. This gives a total $\dot{M}_{\rm ion}$ of $2\,\mathrm{M_{\odot}\,yr^{-1}}$. The outflow momentum flux $\dot{P}_{\rm ion}$ and outflow kinetic luminosity $L_{\rm ion}^{\text{kinetic}}$ are calculated from
\begin{equation}
    \dot{P}_{\rm ion}^{\text {outflow }}=\dot{M}_{\rm ion}\times v_{\rm ion}
\end{equation}
\begin{equation}
L_{\rm ion}^{\text{kinetic}}=\frac{1}{2} \dot{M}_{\rm ion}\times v_{\rm ion}^2.
\end{equation}
We calculate values for $\dot{P}_{\rm ion}^{\text {outflow }}$ and $L_{\rm ion}^{\text{kinetic}}$ of $1.5\times10^{34}\,\mathrm{dyne}$ and $8.0\times10^{41}\,\mathrm{erg\,s^{-1}}$, respectively. These are much smaller ($\sim2\,$dex) than the values derived on larger scales by SINFONI \citep[][]{Nesvadba_2017a,Nesvadba_2017b}. If we use the infrared AGN luminosity, $L_{\rm AGN}^{\rm IR}=10^{10.91}\,L_{\odot}$, and assume a conversion factor, $\kappa_{\rm AGN}^{\rm bol}=6$, the  bolometric luminosity of 4C+19.71 is estimated to be $L_{\rm bol}\sim2.5\times10^{47}\,\mathrm{erg\,s^{-1}}$ \citep[][]{Drouart_2014,Falkendal_2019}. Then the coupling efficiency of this potential radiatively-driven outflow, $L_{\rm ion}^{\text{kinetic}}/L_{\rm bol}$, is $\sim 10^{-5}$. Using NIRSpec IFU, \cite{Perna_2023} and \cite{Vayner_2023Q3D_2} reported higher coupling efficiency for $z\sim3$ quasars ($0.02$ and $10^{-3}$, respectively). \citet{Veilleux_2023} found a relatively low coupling efficiency using the same instrument, $1.8\times10^{-4}$, for a $z\sim1.6$ type-1 quasar which is still 1 dex higher than 4C+19.71. \cite{Harrison_2018} summarised the coupling efficiency for a large number of AGN (e.g., their Fig. 2). Compared to their statistics, the value from this work is located at the lower end. Our estimation based on the \textit{JWST} observation is a $\sim2$ to $3\,$dex lower than the coupling efficiency of the jet kinetic energy probed on larger scales for a  sample of $\sim50$ HzRGs \citep[$10^{-3}$ to $10^{-2}$, ][]{Nesvadba_2008,Nesvadba_2007a,Nesvadba_2017a,Nesvadba_2017b}.  

Our NIRSpec IFU data is only sensitive to the central clumps, $\lesssim20\,$kpc, but not to extended diffuse emission. If we assume there are fast outflows ($\sim500\,\mathrm{km\,s^{-1}}$) and use the distance from the centre quasar to the edge of our FoV ($\sim20\,$kpc), the crossing time of this outflow would be $\sim40\,$Myr. This is shorter than the jet age, $\sim60\,$Myr, of 4C+19.71 \citep[][]{Nesvadba_2017b}. Hence, it is plausible that we do not capture the fast and (presumably) energetic outflowing gas on large scales (e.g., size of the radio lobes). Nevertheless, if the radiative coupling efficiency is $10^{-5}$ at the nucleus where the radiation is the strongest, it will be less efficient farther outside. Simulations \citep[e.g.,][]{Costa_2018a} find that the efficiency for the radiation-driven outflows can be $\sim0.1\%$ in luminous quasars ($L_{\rm bol}\gtrsim10^{47}\,\mathrm{erg\,s^{-1}}$) within galactic scales ($\lesssim10\,$kpc). The efficiency drops after $\sim10\,$Myr which may resemble the case of our observation. Though the ISM and CGM gas of 4C+19.71 seem more quiescent than in other HzRGs \citep[][]{Nesvadba_2017b,wang2023}, a gas kinetic energy luminosity of $10^{45}\,\mathrm{erg\,s^{-1}}$ was found for this source by including the ISM at the jet lobes. Comparing to the results from \citet{Nesvadba_2017a,Nesvadba_2017b}, we conclude that, at least for this HzRG, the mechanical feedback from the radio jet takes the leading role for driving the outflow, and it is happening in the radio lobes beyond the galactic scale (e.g., $\sim$30 kpc from the AGN). 

One of the main uncertainties is the estimation of $M_{\text {ion}}$, which is inversely proportional to $n_{e}$. We find that $M_{\text {ion}}$ (and thus $L_{\rm ion}^{\rm kinetic}$) could be $\sim2\,$dex lower if we use the $n_{e}$ estimated based on our observation (Appendix \ref{app:linelist}). Therefore, the radiative coupling could be even more inefficient. Another uncertainty is the outflow speed. We take the unknown orientation of the ionization cone into account by using $v_{\rm ion} = |\Delta v|+\sigma_{v}$. Even if the estimated speed is one of the three components in 3D space (i.e., the intrinsic $v_{\rm ion}$ is higher by another factor of three), the radiative coupling efficiency will increase by a factor of nine, $\sim10^{-4}$, which is still $\sim1\,$dex lower than the kinetic efficiency from \citet{Nesvadba_2017b}.

\subsection{AGN illuminates the filamentary ISM after jet passage}\label{subsec:view_ISM}

In Sect. \ref{subsec:O3morphology}, we found that ionized ISM emission with $\lesssim30\,$kpc of the AGN is dominated by gas close to the systemic redshift (Fig. \ref{fig:IFU+spec}c). Due to the limitation of the FoV, we only captured the northern extended region. In general, the gas morphology is elongated along the north-south direction, aligned with the jet axis. When we focus on the smaller scale structures ($\sim 5$\,kpc), these gaseous clouds are filamentary and patchy and do not follow the collimated jet but have larger angle separations ($\sim 30\deg$). As discussed in Sect. \ref{subsec:linera_diag}, we are observing the ionization cone-like structures along the jet. This provides enough ionizing photons to illuminate the extended gaseous nebula beyond the galactic scale. The fact that we do not observe the extended ISM in directions outside of the ionization cone, for example in the west where we have the spatial coverage (Fig. \ref{fig:IFU+spec}), may be due to the shortage of ionizing photons.  Hence, it may be that the patchy filaments in the North are the ``intrinsic'' structures, which are not seen in other directions as they are not emitting or simply outside the NIRSpec FoV. 

The HzRGs we observed may have entered their flux-limited parent radio surveys because their radio lobes are brightened when they encounter a higher density environment which provides a larger ``working surface'' for the jet \citep[][]{Eales_1992,wang2023}. Combining the discussion in Sect. \ref{subsec:ouflow} and \citet{Nesvadba_2017b}, the situation of 4C+19.71 may be that the jet was launched $\sim60\,$Myr ago, and in its younger years could have pushed some gas outside of the host galaxy. Once the radio lobes have escaped the host galaxy, the radio jet is very collimated and is no longer interacting with the ISM; the observed systemic gas in Fig.~\ref{fig:IFU+spec} is thus no longer kinematically affected by the radio jets, but the gas is still photo-ionized by the AGN photons within the ionization cone along the jet axis. 

\citet{Fu_2009} found similar quiescent nebulae whose morphology do not fully agree with their radio maps in low-$z$ quasars on comparable physical scales. They proposed that the expanding bubbles following the jet disturb the ISM. 4C+19.71 has one of the most powerful radio jets, $P_{\rm jet}\sim10^{47}\,\mathrm{erg\,s^{-1}}$ \citep[using $\log (P_{1.4\,\rm GHz}/\mathrm{W\,{Hz}^{-1}}) = 28.6\pm0.1$,][]{Nesvadba_2017b,Cavagnolo_2010}. The situation may be different than the low-$z$ low radio power quasars. Nevertheless, we find evidence of shock ionization on the extended northern nebula of which the jet could be the driver (Sect. \ref{subsec:linera_diag}).  Further analysis (e.g., $T_{e}$ mapping) will be helpful to determine the mechanisms.


%

\subsection{Complex southern ISM}\label{subsec:southISM}
In Sect. \ref{subsec:ouflow}, our discussion of the outflow is based on the northern region near the quasar as the nature of the southern ``high-velocity" region is more complicated. In Fig. \ref{fig:IFU+spec}(c2), we clearly find that there are three velocity components with the ``high-velocity" ones (blue and red) dominating the flux. The morphology of the blue and red parts are not spatially overlapping (Fig. \ref{fig:IFU+spec}a, (b1) and (b3)). Such kinematics resemble that of a rotating disk. If we simply take the velocity shifts of the blue and red components from the one spaxel spectrum (Fig. \ref{fig:IFU+spec}(c2)), $\Delta v_{\rm blue}=-585\,\mathrm{km\,s^{-1}}$ and $\Delta v_{\rm red}=535\,\mathrm{km\,s^{-1}}$. This is very high for a rotating disk; using the half size of $\sim4\,$kpc, this would imply a dynamical mass of $\sim2\times 10^{11}\,\mathrm{M_{\odot}}$. We do not detect any continuum emitting source at this position, though we note that the partially overlapping foreground G2 may obscure some of the emission (Fig. \ref{fig:IFU+spec}a). Extracting from a $r=0.1\arcsec$ aperture, the observed flux density upper limit is $\sim10^{-20}\,\mathrm{erg\,s^{-1}\,cm^{-2}\,\mathring{A}^{-1}}$ at $2.5\,\mu$m (i.e., rest frame $V$ band). This then gives an upper limit to the absolute $V$ band magnitude of $\sim -17$ which could be a galaxy with $M_{\star}\sim 10^{8}$ to $ 10^{9}\,\mathrm{M_{\odot}}$ \citep[e.g.,][]{Weaver_2022}. This corresponds to a $M_{\star}/M_{\rm dyn.}\sim10^{-3}$ which is very  small for a galaxy at $z\sim3.5$. The southern region shows lower $\mathrm{[\ion{O}{i}]/H\alpha}$ and $\mathrm{[\ion{S}{ii}]/H\alpha}$ than the rest of the nebula. It also has the potential to have sub-solar metallicity (Sect. \ref{subsec:linera_diag}). Though our estimates have large uncertainties, we cannot exclude the possibility of it being a late stage merger. A more detailed kinematic study is required to unveil the nature of this region with complex kinematics.

\subsection{Summary and future}
In this work, we present the first NIRSpec IFU view of the ionized ISM in the vicinity of the $z\simeq3.59$ radio-loud AGN 4C+19.71. With unprecedented resolution and sensitivity, we study the gas morphology, ionization states and preliminary kinematics within $10\,$kpc of the AGN. This makes it possible to study the role of the jet and radiative feedback near Cosmic Noon and to compare to low-$z$ AGN and theoretical models. We find that the radiation from the hidden quasar dominates the ionization of the ISM of 4C+19.71 which resembles the scenario of other powerful quasars at Cosmic Noon. The radiatively-driven outflow is only found within $5\,$kpc of the AGN and is inefficiently ($\sim 10^{-5}$) coupled to the central  $L_{\rm bol}\sim10^{47}\,\mathrm{erg\,s^{-1}}$ quasar even at the nucleus. Combining with ground-based studies, we conclude that the jet kinetic energy takes a leading role in the feedback of this HzRG at $z\sim3.5$. Line ratio diagnostics infer the existence of a ionization cone along the jet axis. The AGN ionizing photons illuminate the filamentary extended ISM along the cone. The observed morphology may be the ``intrinsic'' shape of the dense gas around this relatively massive galaxy. These conclusions are based on the first source from our sample. The picture of the feedback from the most powerful radio sources will be clearer with our full sample which will allow kinematics studies of the ionized ISM of radio-loud AGN with very different jet morphologies.

\begin{acknowledgements}

This work uses the NASA's Astrophysics Data System and a number of open source software other than the aforementioned ones such as Jupyter notebook \citep[][]{kluyver2016jupyter}; \texttt{matplotlib} \citep[][]{Hunter_2007}; \texttt{SciPy} \citep[][]{virtanen2020scipy}; \texttt{NumPy} \citep[][]{harris2020numpy}; \texttt{Astropy} \citep[][]{Astropy_2018}; \texttt{LMFIT} \citep[][]{newville2016}.

We thank Daizhong Liu and Markus Hundertmark for the useful discussion of the WCS alignment. This work is based in part on observations made with the NASA/ESA/CSA James Webb Space Telescope. The data were obtained from the Mikulski Archive for Space Telescopes at the Space Telescope Science Institute, which is operated by the Association of Universities for Research in Astronomy, Inc., under NASA contract NAS 5-03127 for JWST. These observations are associated with program JWST-GO-01970. Support for program JWST-GO-01970 was provided by NASA through a grant from the Space Telescope Science Institute, which is operated by the Association of Universities for Research in Astronomy, Inc., under NASA contract NAS 5-03127. NLZ and AV are supported in part by NASA through STScI grants JWST-ERS-01335 and JWST-GO-01970. The work of DS was carried out at the Jet Propulsion Laboratory, California Institute of Technology, under a contract with NASA. This paper makes use of the following ALMA data: ADS/JAO.ALMA\#2021.1.00576.S. ALMA is a partnership of ESO (representing its member states), NSF (USA) and NINS (Japan), together with NRC (Canada), MOST and ASIAA (Taiwan), and KASI (Republic of Korea), in cooperation with the Republic of Chile. The Joint ALMA Observatory is operated by ESO, AUI/NRAO and NAOJ.
\end{acknowledgements}

\bibliographystyle{aa}
\bibliography{references}

\begin{appendix}
\section{Astrometric correction}\label{app:astrometry}
\begin{figure*}
    \centering
    \includegraphics[width=\textwidth,clip]{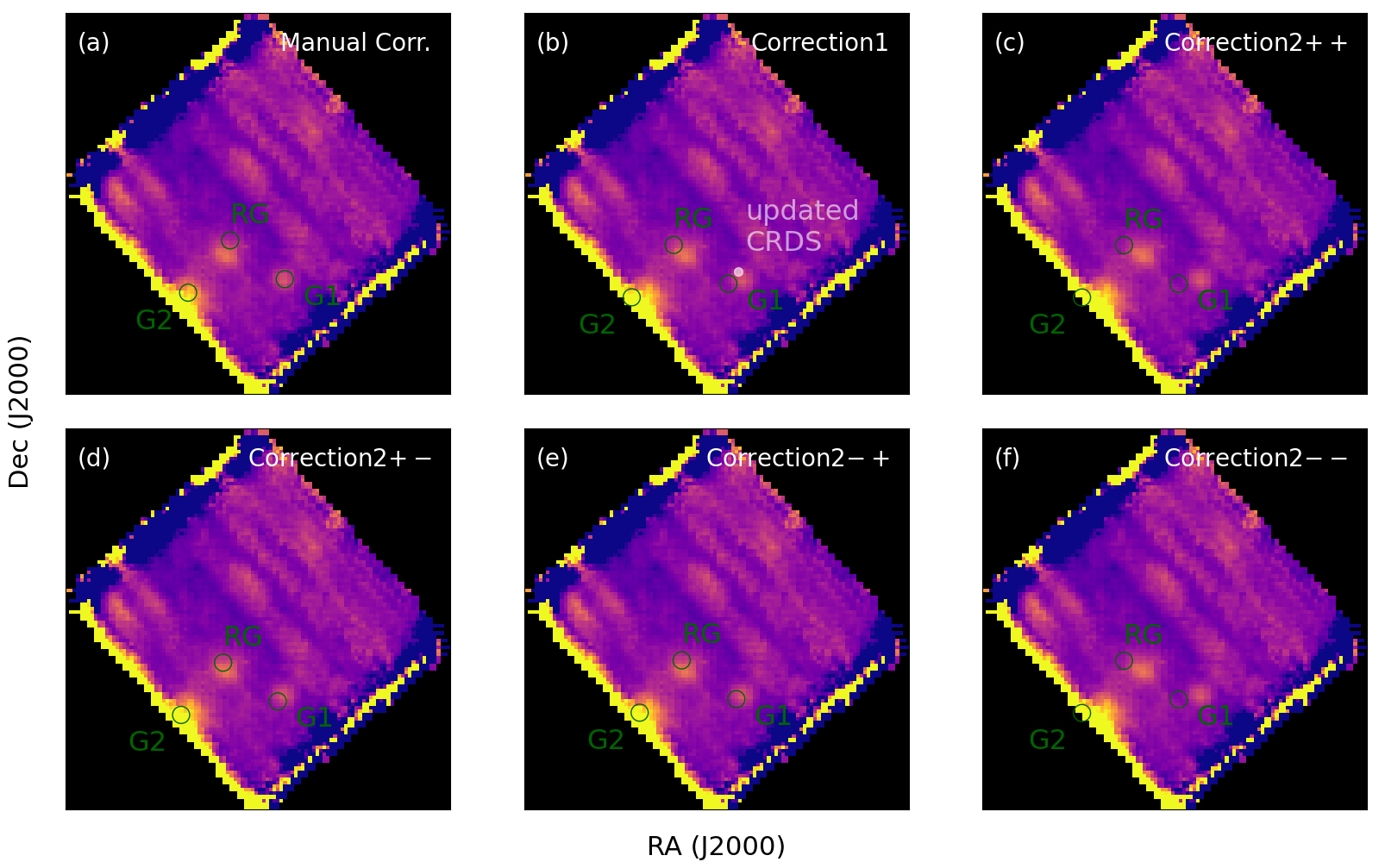}
    \caption{Narrow-band continuum images (observed frame $2.721-2.778\,\mathrm{\mu m}$) with WCS corrected with different methods. The method is labeled at the top right corner of each panel. The three green circles mark the positions of the host galaxy of 4C+19.71 (RG), G1 and G2. The radius of each circle is $0.12\arcsec$ which can be used roughly as the uncertainty of the \textit{HST} WCS. The white dot in panel (b) marks the centroid of G1 using the updated CRDS (jwst\_1093.pmap).}
    \label{fig:wcs_corr}
\end{figure*}

The \textit{JWST} Fine Guidance Sensor (FGS) is responsible for guiding the pointing, and there is no known issue in the guiding process. However, there is a known systematic WCS offset of the NIRSpec IFU\footnote{\url{https://jwst-docs.stsci.edu/jwst-observatory-characteristics/jwst-pointing-performance}}. This is due to poorly registered IFU detector coordinates on the \textit{JWST} focal panel with respect to the reference V2 and V3 axes. This corresponds to 0.173 \arcsec and 0.178 \arcsec in the V2 and V3 direction\footnote{\url{https://jwst-docs.stsci.edu/jwst-observatory-characteristics/jwst-field-of-view}}, respectively (Help Desk, private communication through). This offset projection to sky (RA, Dec) coordinates could then be calculated using \texttt{pysiaf} by assuming the targeted RA and Dec of the IFU pointing is correct. The projection matrix for converting V2, V3 coordinates to the sky plane can be set by \texttt{pysiaf.utils.rotations.attitude\_matrix} using the old (V2, V3) reference coordinates, target RA and Dec, and angle between the north and V3. This angle is measured from north to east with a value of $83.0652845\deg$ for our observation. It is calculated using the angle between the IFU with respect to the North (from the header) and the fixed angle between the IFU and V3. There is an additional $\sim3\arcsec$ systematic rotational angle error between the IFU plane and V3 axis. Since this will only result in a secondary ($\sim0.007\arcsec$) offset, we neglect it here. The final systemic offsets of the pointing, from updated position to the original position, are $\Delta$RA$=0.21\arcsec$ and $\Delta$Dec$=-0.15\arcsec$. We refer to this correction as ``Correction1". This error has been reported as partially corrected using the up-to-date context file. We test the new pipeline assigned WCS using jwst\_1093.pmap and find it matches closely with our ``Correction1". We report an offset of $|\Delta$RA$|=0.09\arcsec$ and $|\Delta$Dec$|=0.07\arcsec$ using the fitted 2D Gaussian centroids of G1 at $\sim2.7\rm \mu m$. Hence, we consider this systematic WCS error has been fixed. However, other source of uncertainty is still presented and still required manual alignment.

In addition, there is a further offset due to the zero-point shift of the NIRSpec detector and the FGS which adds another term to the aforementioned values in the V2V3 plane. The shifts are -0.016\arcsec and 0.111\arcsec in V2 and V3 coordinates, respectively (Help Desk, private communication). However, a caution is noted that the direction of this shift is not well-understood. We assumed this direction and followed the aforementioned procedure using \texttt{pysiaf}. This resulted in the $\Delta$RA$=0.09\arcsec$ and $\Delta$Dec$=-0.14\arcsec$. We refer to this correction as ``Correction2$++$". Since the direction is uncertain, we produced further corrections ``Correction2$+-$", ``Correction2$-+$" and ``Correction2$--$". The ``$+$" and ``$-$" signs indicate which direction we adopted for the offset with respect to the original direction suggested by, the Help Desk.

For the manual correction of the astrometry, we matched G1 positions (Section \ref{sec:results}) measured from the $\sim 2.7\rm \mu m$ continuum image collapsed from our NIRSpec IFU cube and the \textit{HST}/WFPC2 \citep[][]{Whitmore_2016} image. This was done by matching the centroids of the fitted 2D Gaussian profile. We caution that although we updated the astrometry of the \textit{HST} image using \textit{Gaia} DR3 \citep[][]{gaiaDR3_2023}, the uncertainty of the WCS position is $\sim0.06\arcsec$. This is estimated based on the standard deviation of the offsets of the cross matched \textit{Gaia} targets in the \textit{HST} FoV. This uncertainty ($\sim1.2\arcsec$ given the unknown direction) is marked by the size of the plus marker in Fig. \ref{fig:IFU+spec}d. We further note that the \textit{HST} image was observed in 1997. The proper monition of the stars were taken into account during the correction by adopting the reported epoch$=$J2000 coordinates in the \textit{Gaia} DR3 catalog. Another source of uncertainty could be the intrinsic continuum emission offset from different bands. The \textit{HST}/WFPC2 image was taken with the F702W filter, which corresponds to $2520\,\mathrm{\AA}$ in the rest frame of G1. The NIRSpec IFU data covers $\sim 0.57-1.15\mathrm{\mu m}$ for G1. We examined the continuum centroid offset of G1 by extracting narrow-band images at the blue and red ends of the data cube. A $\sim0.09\arcsec$ shift is seen.

We present the offsets of these corrections with respect to the manual shift (aligned with \textit{HST}) used in the analysis in Table \ref{tab:WCS_offset}. Fig. \ref{fig:wcs_corr} presents the visual inspection of the WCS correction by comparing the results from different methods. We mark the \textit{HST} positions of the three continuum emission objects with green circles. The radius of each circle is $0.12\arcsec$ which can be used as the spatial uncertainty of the \textit{HST} object. We see that Correction2$+-$ and Correction2$-+$ have $\sim0.1\arcsec$ offsets from the manual shift (or \textit{HST} WCS). Given that there is a $\sim0.1\arcsec$ uncertainty in the \textit{HST}-NIRSpec alignment, we decide that the IFU WCS offset can be solved by correcting the systematic error. Since the direction of the error is not well-understood, we use the manual match in this paper.
\begin{table}[]
    \centering
        \caption{Offset between different astrometry and manual shift.}
    \label{tab:WCS_offset}
    \begin{tabular}{lcc}

    \hline
    \hline
    Method & $\Delta \alpha_{\rm manual}$ &
     $\Delta \delta_{\rm manual}$\\
     \hline
    Correction1   & 0.22$\arcsec$  & -0.06$\arcsec$ \\
    Correction2$++$ & -0.34$\arcsec$  & 0.07$\arcsec$ \\
    Correction2$+-$ & -0.10$\arcsec$  & 0.09$\arcsec$ \\
    Correction2$-+$ & -0.10$\arcsec$  & 0.06$\arcsec$ \\
    Correction2$--$ & -0.34$\arcsec$  & 0.07$\arcsec$ \\
    \hline
    \end{tabular}

\end{table}

\section{Foreground galaxies}\label{app:fore_galaxy}
\begin{figure}[h!tbp]
    \centering
    \includegraphics[width=0.5\textwidth,clip]{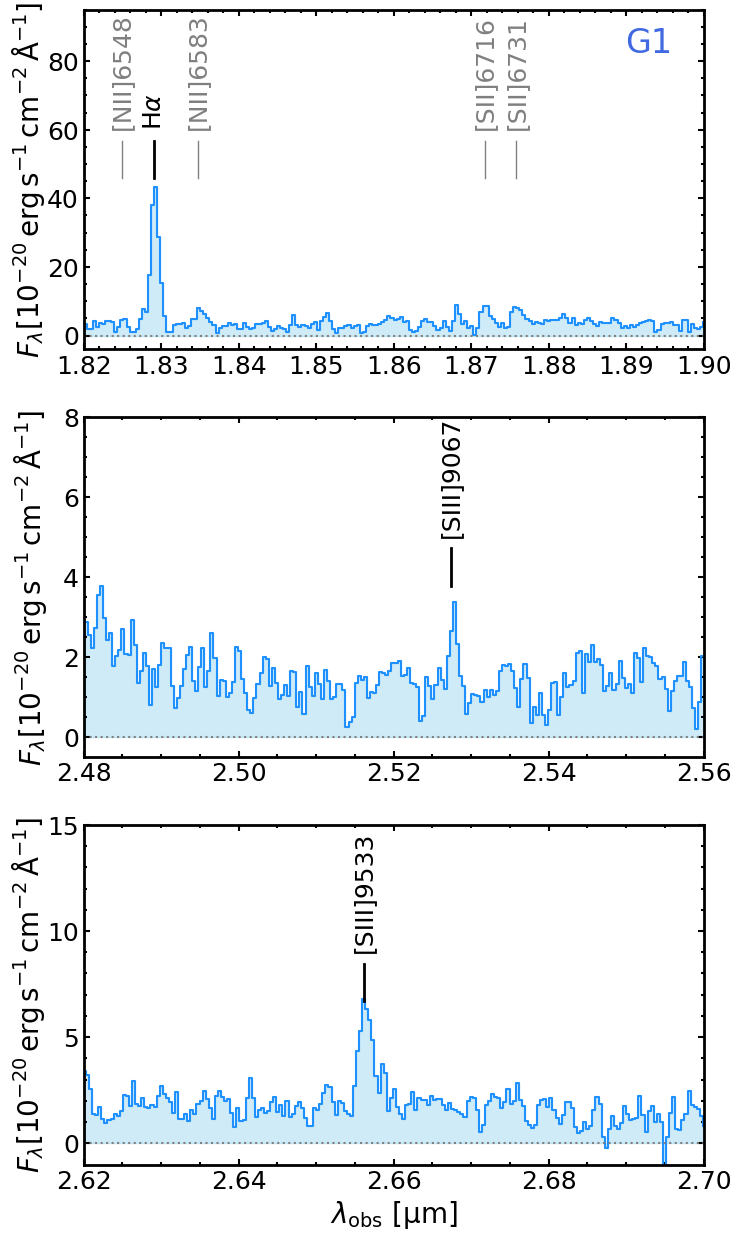}
    \caption{Spectrum extracted at the position of foreground galaxy, G1, without continuum subtraction. \textbf{Upper panel:} Zoom-in view of the spectrum around H$\alpha$. \textbf{Middle panel:} Zoom-in view of the spectrum around [\ion{S}{iii}]9071. \textbf{Bottom panel:} Zoom-in view of the spectrum around [\ion{S}{iii}]9533. We mark the lines detected both spectrally and spatially in thick black lines. The expected positions of the lines around H$\alpha$ are marked in grey lines. }
    \label{fig:spec_G1}
\end{figure}

\begin{figure}[h!tbp]
    \centering
    \includegraphics[width=0.5\textwidth,clip]{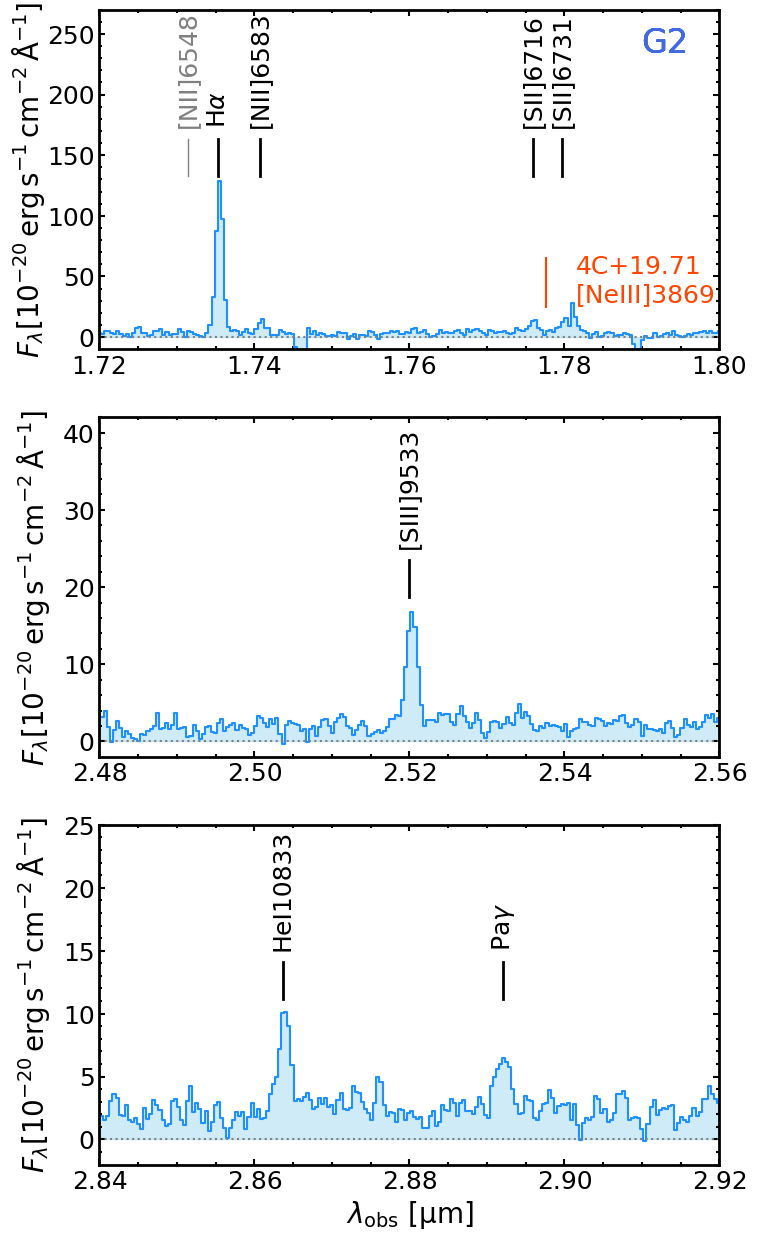}
    \caption{Spectrum extracted at the position of foreground galaxy, G2, without continuum subtraction. \textbf{Upper panel:} Zoom-in view of the spectrum around H$\alpha$. \textbf{Middle panel:} Zoom-in view of the spectrum around [\ion{S}{iii}]9533. \textbf{Bottom panel:} Zoom-in view of the spectrum around \ion{He}{i}10833. Same line-marking conventions are adopted as Fig. \ref{fig:spec_G1}. We further mark in orange the position of [\ion{Ne}{iii}]3869 emission from 4C+19.71 which is near the [\ion{S}{ii}]$6716,6731$ doublet of G2 (upper panel).}
    \label{fig:spec_G2}
\end{figure}

We reported in Sect. \ref{sec:results} that two foreground galaxies have been identified with continuum and emission lines: G1 to the west of the radio galaxy and G2 to the east. There we present the spectra of G1 and G2. Their positions are marked with blue dots in Fig.\ref{fig:IFU+spec} which are the centers of their extraction apertures. For both galaxies, we used square apertures with the size of $0.25\arcsec\times0.25\arcsec$ ($\sim5\times5$ pix$^{2}$) and summed each spectrum from the spaxels inside the aperture. For partially covered spaxels, we multiply the area by the fractional overlap with the extraction aperture. The aperture for G1 was centered at its continuum peak position. As for G2 which is only partially captured in our FoV, we placed the aperture to be as close as possible to its continuum emission peak while minimising the impact from the detector edge. In Fig. \ref{fig:spec_G1} and \ref{fig:spec_G2}, we show the spectra extracted for G1 and G2, respectively. The redshifts of G1 and G2 are determined based on their H$\alpha$ lines and have values of $z_{\rm G1} \simeq 1.786$ and $z_{\rm G2} \simeq 1.643$. For each foreground galaxy, at least two more emission lines are identified based on their redshifts. We use thick black vertical lines to mark the spatially and spectrally detected emission lines. The expected positions of emission lines near H$\alpha$ based on the derived redshift of the targets are marked in grey lines if not clearly detected. We note that the [\ion{S}{ii}]$6716,6731$ doublet of G2 overlaps with the [\ion{Ne}{iii}]3869 line of 4C+19.71 in wavelength (orange line in top panel of Fig. \ref{fig:spec_G2}). The spatial check confirms that there are distinct emission peaks separated by $\sim0.4\arcsec$. 

\section{Emission line list}\label{app:linelist}

\begin{table}[]
    \centering
        \caption{Fitted emission line fluxes. The reported uncertainty is the direct $1\sigma$ output of \texttt{q3dfit}.}
    \label{tab:emissionline}
    \begin{tabular}{lccc}

    \hline
    \hline

    Line & central comp. & red comp. &  blue comp. \\
    & \multicolumn{3}{c}{$f\, [10^{-18}\,\mathrm{erg\,s^{-1}\,cm^{-2}}$]} \\
     \hline
    $[\ion{O}{ii}]3726,3729$                & 5.8$\pm$0.4  & 2.1$\pm$0.4 & 8.9$\pm$1.6 \\
    $[\ion{Ne}{iii}]3869$                   & 3.3$\pm$0.2  & 0.8$\pm$0.2 & 1.3$\pm$0.9 \\
    $\ion{He}{i}3888+\mathrm{H8}^{*}$       & \multicolumn{3}{c}{$<0.5$}  \\
    $[\ion{Ne}{iii}]3967+\mathrm{H\epsilon}$& 1.5$\pm$0.4  & 0.4$\pm$0.3 & 0.4$\pm$1.4\\
    H$\delta$                               & 0.7$\pm$0.3  & 0.2$\pm$0.2 & 0.6$\pm$0.9\\
    H$\gamma$                               & 1.6$\pm$0.2  & 0.4$\pm$0.2 & 0.4$\pm$0.7\\
    $\ion{He}{i}4472^{*}$                   & \multicolumn{3}{c}{$<0.4$}  \\
    $\ion{He}{ii}4686$                      & 1.4$\pm$0.1  & $<0.2^{\dag}$      & 1.1$\pm$0.8\\
    H$\beta$                                & 3.9$\pm$0.2  & 0.6$\pm$0.2 & 2.9$\pm$0.7\\
    $[\ion{O}{iii}]4959$                    & 17.5$\pm$0.5 & 1.8$\pm$0.4 & 8.7$\pm$1.5\\
    $[\ion{O}{iii}]5007$                    & 53.1$\pm$1.0 & 5.6$\pm$1.0 & 26.3$\pm$1.6\\
    $\ion{He}{i}5876^{*}$                   & \multicolumn{3}{c}{$<0.6$}  \\
    $[\ion{Fe}{vii}]6087^{*}$               & \multicolumn{3}{c}{$<0.7$}  \\
    $[\ion{O}{i}]6300$                      & 2.4$\pm$0.2  & 0.3$\pm$0.2& 3.1$\pm$0.8\\
    $[\ion{S}{iii}]6312$                    & 0.4$\pm$0.2  & 0.2$\pm$0.2& 0.9$\pm$0.9\\
    $[\ion{O}{i}]6364$                      & 0.8$\pm$0.3  & 0.1$\pm$0.3& 1.0$\pm$1.1\\
    $[\ion{N}{ii}]6548$                     & 2.0$\pm$0.3  & 0.7$\pm$0.3& 4.0$\pm$1.1\\
    H$\alpha$                               & 15.8$\pm$0.4 & 1.2$\pm$0.3& 14.7$\pm$1.1\\
    $[\ion{N}{ii}]6583$                     & 6.0$\pm$0.4  & 2.0$\pm$0.3& 12.0$\pm$1.1\\
    $[\ion{S}{ii}]6716$                     & 1.6$\pm$0.2  & 0.3$\pm$0.2& 2.1$\pm$0.6\\
    $[\ion{S}{ii}]6731$                     & 1.8$\pm$0.2  & 0.6$\pm$0.2& 4.8$\pm$1.1\\
    
    \hline
    \end{tabular}
    \tablefoot{$^{*}$ We report $3\sigma$ upper limits for lines with low S/N (Sect. \ref{subsec:line_emission}). $^{\dag}$  Though the $\ion{He}{ii}4686$ has $S/N>10$, the fit reported the flux of its comp. red to be $0$. Hence, we report the upper limit here.}

\end{table}

We fit the 1D spectrum presented in Section \ref{subsec:line_emission} (Fig. \ref{fig:spec_full}) using \texttt{q3dfit}. The continuum is modeled by a third order polynomial and the emission lines (S/N$>10$) are fitted with Gaussian profiles. We set the maximum number of the Gaussian components to be three for the fit. The centroid (width) of each corresponding Gaussian component is set to be the same \citep[][]{Vayner_2023Q3D}. The fitted line fluxes are shown in Table \ref{tab:emissionline}. For the doublet $[\ion{O}{ii}]3726,3729$, we set the flux ratio to be $1:1$ and report the summation since they are spectrally unresolved. For the low S/N lines (Section \ref{subsec:line_emission}), we report $3\sigma$ upper limits.

\section{Systemic redshift based on ionized gas}\label{app:sys_z}

The systemic redshift (velocity zero) used in this work is based on the [\ion{C}{i}](1-0) \citep[z=3.5892,][]{kolwa2023}. Given its relatively faint flux and large observed beam size, it is unclear whether this traces the ``authentic'' redshift of the host galaxy. The ideal tracer would be the stellar absorption lines (e.g., \ion{Ca}{ii} HK). However, the continuum emission near the nucleus is AGN dominated (Sect. \ref{subsec:low_z_comp}) for our target. Nevertheless, we report the [\ion{O}{iii}] flux-weighted redshift based on the fitted ``low-velocity'' components from the entire FoV (Sect. \ref{subsec:linera_diag}). We define the ``low-velocity'' components as the ones with $|\Delta v|<300\,\mathrm{km\,s^{-1}}$ \citep[e.g.,][]{Fu_2009}. The redshift is $z_{[\ion{O}{III}]}=3.5908\pm0.0017$ which corresponds to $108\pm114\,\mathrm{km\,s^{-1}}$ with respect to the [\ion{C}{i}](1-0) redshift. The  $z_{[\ion{O}{III}]}$ and $z_{[\ion{C}{i}]}$ are consistent within the relatively large uncertainty. 

In our calculation in Sect. \ref{subsec:ouflow}, we assume the central component is the systemic component. Its $\Delta v_{\rm central}=238\pm10\,\mathrm{km\,s^{-1}}$ agrees with the systemic velocity estimated here from the entire FoV.

\section{Estimation of electron density}\label{app:n_e}

We use the fiducial value of $n_{e}$ in the calculation in Sect. \ref{subsec:ouflow} \citep[][]{Nesvadba_2017b}. The $n_{e}$ for the red and blue components can also be estimated based on the fitted [\ion{S}{ii}]6716 and [\ion{S}{ii}]6731 lines ratios (0.50 and 0.44 for red and blue components respectively; Table \ref{tab:emissionline}). This result in a $n_{e}\,\sim8\times10^{3}\mathrm{cm^{-3}}$ and $\sim8\times10^{4}\mathrm{cm^{-3}}$ for the red and blue components, respectively \citep[e.g,][]{Sanders_2016}. These are much larger than the fiducial value, $n_{e}=500\,\mathrm{cm^{-3}}$, from \citet{Nesvadba_2017b}. However, the fitted line ratios for these two kinematic components are in the range where they are not sensitive to $n_{e}$. If we nevertheless use $n_{e}$ based on our estimation, the radiative coupling efficiency becomes $\sim10^{-7}$, and continues to support the conclusion is valid that the radiative-driven outflow is inefficient in 4C+19.71.

 We report that the $n_{e}$ of the central component is $\sim800\,\mathrm{cm^{-3}}$ based on its [\ion{S}{ii}]6716 and [\ion{S}{ii}]6731 ratio \citep[0.89, Table \ref{tab:emissionline};][]{Sanders_2016} which is the same order of magnitude as the value reported by \citet{Nesvadba_2017b}. Further analysis is required to understand the electron temperature and density at the center.

\end{appendix}
\end{document}